\documentclass[twocolumn,prb,amsmath,amssymb,superscriptaddress]{revtex4-1}
\usepackage{graphicx,bm}
\usepackage{hyperref}
\hypersetup{colorlinks=true,linkcolor={black}}

\def\mbF{{\mathbf F}}

\def\<{{\langle}}
\def\>{{\rangle}}

\newcommand{\al}{\alpha}
\newcommand{\be}{\beta}

\newcommand{\de}{\delta}
\newcommand{\D}{\Delta}
\newcommand{\e}{\epsilon}

\newcommand{\G}{\Gamma}
\newcommand{\kap}{\kappa}
\newcommand{\lm}{\lambda}
\newcommand{\Lm}{\Lambda}
\newcommand{\w}{\omega}
\newcommand{\W}{\Omega}
\newcommand{\z}{\zeta}

\newcommand{\s}{\sigma}
\newcommand{\st}{\sigma\tau}

\newcommand{\del}{\nabla}

\newcommand{\p}{\partial}

\newcommand{\delp}{\nabla_\mathbf{p}}

\newcommand{\bsdel}{\boldsymbol{\nabla}}

\newcommand{\vphi}{\vec{\phi}}

\newcommand{\mbj}{\mathbf{j}}

\newcommand{\hatbp}{\hat{\mathbf{p}}}
\newcommand{\mbp}{\mathbf{p}}

\newcommand{\mbps}{{\mathbf{p}\sigma}}

\newcommand{\mbP}{\mathbf{P}}
\newcommand{\mbq}{\mathbf{q}}

\newcommand{\mbr}{\mathbf{r}}

\newcommand{\mbv}{\mathbf{v}}

\newcommand{\bsphi}[1]{\boldsymbol{\phi}}
\newcommand{\mbf}[1]{\mathbf{#1}}

\newcommand{\bfhat}[1]{\hat{\mathbf{{#1}}}}

\newcommand{\mcal}[1]{\mathcal{#1}}

\newcommand{\til}[1]{\tilde{#1}}
\newcommand{\coltwo}[2]{\left(\begin{array}{c}#1\\#2\end{array}\right)}

\newcommand{\colfour}[4]{\left(\begin{array}{c}{#1}\\{#2}\\{#3}\\{#4}\end{array}\right)}
\newcommand{\pfrac}[2]{\left(\frac{#1}{#2}\right)}

\newcommand{\hatC}{\hat{C}}

\newcommand{\bs}[1]{\boldsymbol{#1}}

\newcommand{\nn}{\nonumber\\}

\newcommand{\ben}{\begin{equation}}
\newcommand{\een}{\end{equation}}

\newcommand{\app}[1]{Appendix \ref{#1} }

\newcommand{\intdppp}{\int{ d\mbp\over(2\pi\hbar)^3}}

\newcommand{\intp}{\int{ d\mbp\over(2\pi\hbar)^3}}

\begin{document}
\title{Spin-heat relaxation and thermo-spin diffusion in atomic Bose and Fermi gases}
\author{Clement H. Wong}
\affiliation{Department of Physics, University of Wisconsin-Madison, Madison, Wisconsin 53706, USA}
\author{H.T.C. Stoof}
\affiliation{Institute for Theoretical Physics and and Center for Extreme
Matter and Emergent
Phenomena, Utrecht University, Leuvenlaan 4, 3584 CE Utrecht, The Netherlands}
\author{ R.A. Duine}
\affiliation{Institute for Theoretical Physics and and Center for Extreme
Matter and Emergent
Phenomena, Utrecht University, Leuvenlaan 4, 3584 CE Utrecht, The Netherlands}
\date{\today}
\begin{abstract} 
We study spin-dependent heat transport in quantum gases, focusing on transport phenomena related to pure spin currents and spin-dependent temperatures.   Using the Boltzmann equation, we compute the coupled spin-heat transport coefficients as a function of temperature and interaction strength for energy dependent $s$-wave scattering.   We  address the issue of whether spin-dependent temperatures can be sustained on a time and length scale relevant for experiments by computing the spin-heat relaxation time and diffusion length.  We find that the time scale for spin-heat relaxation time diverges at low temperatures for both bosons and fermions, indicating that the concept of spin-heat accumulation is well defined for degenerate gases.   
For bosons, we find power-law  behavior on approach to Bose condensation above the critical temperature, as expected from the theory of dynamical critical phenomena.   
\end{abstract}
\maketitle
\section{Introduction}

Spin caloritronics is currently an active field of research concerned with studying the spin-dependent generalizations of thermoelectric effects in solid-state materials,\cite{bauerNAT12}  as well as novel collective effects.  Just as the traditional thermoelectric phenomena, i.e., the Seebeck and Peltier effects, have applications in generators, refrigerators, and in utilizing waste heat, thermally driven spin currents may have  applications in spintronics devices.  In fact, the coupling of particle, (pseudo) spin and heat transport is a general phenomenon not restricted to the solid-state environment.   Thermoelectric effects in ultracold atomic gases have recently become a topic of experimental interest.\cite{brantutSCI13,grenierCM12,ranconCM13,HazlettCM13}  In contrast to the multifaceted  mechanism of heat transport in the solid-state, which includes disorder and phonon scattering, electron-electron interactions, and in ferromagnetic materials,  spin-polarized conductivities, magnon scattering, spin-flip scattering,\cite{hatamiSSC10,nunnerPRB11}   in cold atoms, atom-atom interactions are the only natural source of scattering, which can even be controlled experimentally through Feshbach resonances.  Thus, the cold atomic gases provide a clean and controllable environment for studying thermoelectric and spin caloritronic effects at the fundamental level.  Conversely, measurements of the spin-heat transport coefficients can be used to extract information about the scattering processes.

In this paper, we consider two-component (pseudo-) spin $1/2$ atomic gases in a smooth trapping potential, in mechanical equilibrium where the net forces on the cloud are balanced by the trapping forces.\cite{Note0a}
Even for this stationary gas, without any spin polarization, a pure spin current can be established in response to opposite forces on each spin, i.e., a spin force, due to interspin scattering that transfer momentum between opposite spins.  This viscosity between spins is called spin drag and has been calculated and measured in Bose and Fermi gases,\cite{vichiPRA99,sommerNAT11,kollerCM12,poliniPRL07} and its contribution to the spin diffusion coefficient for electrons, called spin Coulomb drag, has been measured in GaAs quantum wells.\cite{weberNAT05}    Due to the Peltier effect, this spin current is accompanied by a spin-heat current, a difference in the heat currents carried by each spin. The thermodynamic reciprocal effect is the spin-Seebeck effect, by which a spin current is driven by gradients of the spin-heat accumulation, i.e., opposite temperature gradients for the two spin states.  This coupling is generic, so that for example, in the experiment of Ref.~[\onlinecite{sommerNAT11}], spin-dependent heating will occur in the presence of spin currents.

A natural question which arises in considering spin-dependent heat transport is whether one can in practice sustain spin-dependent temperatures, which, in the absence of externally applied spin dependent heating, will ultimately equilibrate due to interspin scattering.   In fact, systems that are modeled with multiple temperatures occur in many subfields in physics, including two-component plasmas with large mass differences,\cite{ohdePP96}  magnetic systems excited by femtosecond laser pulses,\cite{sultanPRB12} and in nanopillar spin valves where the difference between spin up and spin down temperatures, called the spin-heat accumulation, and the associated spin-heat relaxation rates and lengths have been measured.\cite{dejeneNATP13,marunPRL14}  In this paper,  we address this issue specifically for the case of ultracold atomic gases.\cite{Note0}    We show that the spin-heat accumulation can be treated as a quasi-equilibrium quantity much like spin accumulation, i.e., spin-dependent chemical potentials, in the presence of spin-flip scattering. We compute the spin-heat relaxation time and length as functions of temperature and interaction strength, and  find power-law divergences for  the relaxation time at degenerate temperatures for both bosons and fermions, indicating that the spin-heat accumulation is in principle well defined for degenerate quantum gases.  We also find that, depending on the interspin scattering lengths, the relaxation length can be on the order of $\mu$m's for bosons and mm's for fermions,  which is well within experimental resolution, and comparable to or larger than the system size. 

Thermally driven spin currents can be utilized in spintronic devices, for example, to move a domain wall.\cite{kovalevSSC10} Similarly, the coupled spin-heat transport we study here may  be utilized for atomtronic devices that run on spin currents.  Therefore, we introduce a dimensionless quantity characterizing spin-heat conversion in this system called ``$Z_sT$" in analogy to the ``$ZT$"  figure of merit that determines the efficiency of solid-state   thermoelectric devices.   We find that for bosons, with strong scattering,   $Z_sT$ and the spin-Seebeck coefficient are enhanced on approach to the critical temperature of Bose-Einstein condensation, in contrast to the case of weak scattering in Ref.~[\onlinecite{wongPRL12}].  At weak scattering, we also find a sign change in the spin-Seebeck coefficient.  

This paper is organized as follows.  In Sec.~\ref{hydro}, we introduce the Boltzmann and associated hydrodynamic equations for a two-component gas. In Sec.~\ref{Ts}, we report the results of our calculations of the spin-heat relaxation times and lengths for bosons and fermions.  In Sec.~\ref{linear}, we express, in linear response, spin-dependent response and relaxation coefficients in terms of the collision integral, specializing to the case of unpolarized gases in Sec.~\ref{unpolarized}. In Sec.~\ref{moment}, we develop a moment expansion for the computation of the collision integrals which explicitly preserves Onsager reciprocity.  In Sec.~\ref{swave}, we present our results for the transport coefficients as a function of temperature and interaction strength for bosons, extending the work of Ref.~[\onlinecite{wongPRL12}] to include dependence on scattering length. Relevant thermodynamic properties are summarized in App.~\ref{sec:thermo}, and computation details are given in App.~\ref{collisionMat}.


\section{spin-dependent Boltzmann and hydrodynamic equations\label{hydro}}
We will compute the transport coefficients of the two-component gas using the semiclassical Boltzmann equation for the distribution functions $n_\mbps(\mbr,t)$, given by 
\ben
(\p_t+\mbv_\mbp\cdot\bsdel+\mbf{f}_\s\cdot\bsdel_\mbp)n_ {\mbps}(\mbr,t)=\mcal{C}_{\mbps}[n_+,n_-]\,,\label{boltz}
\een
where  $\s$=$\pm$ label the pseudospin index, $\mbf{f}_\s$ are external forces, and
\begin{widetext}
\begin{align}
\mcal{C}_{\mbp_1\s} [n_+,n_-]={2\pi\over\hbar}\prod_{i=2}^4\int{d\mbp_i\over(2\pi\hbar)^3}&(2\pi\hbar)^3\de^3(\mbp_1+\mbp_2-\mbp_3-\mbp_4)\de(\e_{\mbp_1}+\e_{\mbp_2}-\e_{\mbp_3}-\e_{\mbp_4})\nn
&\sum_{\tau=\pm} W_{\s\tau} [n_{3 \s}  n_{4 {\tau}}(1+\zeta n_{1 \s})(1+\zeta n_{2 \tau})-n_{1 \s}  n_{2 {\tau}}(1+\zeta n_{3 \s})(1+\zeta n_{4 {\tau}})]\,,
\label{coll}
\end{align}
\end{widetext}
is the collision integral that describes two-body elastic and spin-conserving scattering of particles from the momentum and spin states $(\mbp_1\s,\mbp_2\tau)$ to $(\mbp_3\s,\mbp_4\tau)$, and $\zeta$=$\pm1$ pertains to bosons $(+)$ and fermions $(-)$.  In Eq.~\eqref{coll}, we defined transition probabilities $W_{\st}$ that takes into account Bose and Fermi particle statistics, given by
\begin{align}
W_{+-}(p_r,\chi)&=|T_{+-}(p_r,\chi)|^2=W_{-+}(p_r,\chi)\,,\nn
W_{\s\s}(p_r,\chi)&={1\over2}|T_{\s\s}(p_r,\chi)+\zeta T_{\s\s}(p_r,\pi-\chi)|^2\,,
\label{W}
\end{align}
where ${T_{\s\tau}(p_r,\chi)\equiv\<\mbp_r',\st|\hat{T}|\mbp_r,\st\>}$ is the two-body transition matrix element  between  incoming and outgoing relative momenta $\mbp_r$=$(\mbp_1$$-$$\mbp_2)/2$ and $\mbp_r'$=$(\mbp_3$$-$$\mbp_4)/2$, respectively, $p_r$=$|\mbp_r|$, and $\chi$ is the angle between relative momenta defined by $\cos\chi$$\equiv$$\bfhat{p}_r$$\cdot$$\bfhat{p}'_r$, where $\bfhat{p}=\mbp/|\mbp|$.  
The transition probabilities in Eq.~\eqref{W} are related to the differential cross section for scattering between spin $\s$ and $\tau$ particles by $d\s_{\s\tau}/d\Omega$$=$$({m}/{4\pi\hbar^2})^2W_{\st}$.  While the formalism we present in the following applies for a generic spin-dependent scattering cross section, we will specifically compute transport coefficients for $s$-wave scattering which is independent of $\chi$, 
\[\frac{d\s_{+-}}{d\Omega}=\frac{a^2}{1+(p_ra/\hbar)^2}\,,\]
where $a$ is the interspin $s$-wave scattering length.\cite{Note5a} For bosons, we consider equal interspin and intraspin scattering lengths, so that ${d\s_{\s\s}}/{d\Omega}$=$2{d\s_{+-}}/{d\Omega}$ .

The hydrodynamic equations for the spin $\s$ particle density, momentum, and energy densities  given by taking $\intp\{1,\mbp,\e_\mbp\}\times$  Eq.~\eqref{boltz}, respectively, are
\begin{align}
\p_t\rho_\s+\bsdel\cdot(\rho_\s\mbv_\s)&=0\,,
\label{particle}\\
m\rho_\s(\p_t+\mbv_\s\cdot\bsdel)\mbv_\s-\rho_\s\mbf{f}_\s&=-\bsdel\cdot\tensor{\bm{\pi}}_\s+\bm{\Gamma}_\s\,,
\label{euler}\\
\p_te_\s+\bsdel\cdot\mbj_ {e\s}&=\mbf{f}_\s\cdot\mbj_\s+\Gamma_\s\,,
\label{energy}
\end{align}
where the particle density, particle and energy current, and average velocities densities are defined by
\ben
\colfour{\rho_\s}{e_\s}{\mbj_\s}{\mbj_{e\s}}\equiv \intp\colfour{1}{\e_\mbp}{\mbv_p}{\e_\mbp\mbv_p}n_{\mbp\s}\,,
\label{densities}
\een
where $\mbj_\s$$\equiv$$\rho_\s\mbv_\s$, $\mbv_p$$=$$\delp\e_\mbp$, and the stress tensor is 
\[\pi_{ij\s}\equiv\rho_\s v_{\s i}v_{\s j}-{1\over m}\intp p_{i} p_{j}n_\mbps\,,\]
and we defined 
\ben
\coltwo{\Gamma_\s}{\bm{\Gamma}_\s}=\intp\coltwo{\e_\mbp}{\mbp} \mcal{C}_{\mbps}[\vec{n}]\,,
\label{Gamma}
\een
where  $\vec{n}_\mbp=(n_{\mbp+},n_{\mbp-})$, and henceforth an arrow denotes vectors in spin space.
These source terms represent the transfer of  energy and momentum through inter-spin scattering, and are proportional to the spin drag and spin-dependent temperature relaxation rates, which are the focus of this work.

We note here that the spin $\s$ collision integrals and their sum possess  collisional invariants corresponding to conservation laws.  The particle continuity equation  Eq.~\eqref{particle} reflects the conservation of the spin $\s$ particle number  in the absence of spin-flip scattering, so that 
\[\intp \mcal{C}_{\mbps}[\vec{n}]=0\,,\]
and furthermore, since the total  energy and momentum  is conserved, we must have
\ben
\sum_\s \coltwo{\Gamma_\s[\vec{n}]}{\bm{\Gamma}_\s[\vec{n}]}=0\,.
\label{Cinv}
\een
These identities will be used in the subsequent sections.  
  
Since we will be interested in heating, we transform the energy equation Eq.~\eqref{energy} into an entropy-production equation following standard fluid mechanics,\cite{landauFM} and we find
\ben
\rho_\s T_\s (\p_t+\mbv_\s\cdot\bsdel)s_\s=\mbf{f}_\s\cdot\mbj_\s-\bsdel\cdot\mbq_\s+\Gamma_\s+\mbv_\s\cdot\bm{\G}_\s\,,
\label{entropy}
\een
where $T_\s$ and $s_\s$ is the spin $\s$ temperature and entropy per particle respectively, and we define the heat current by
\ben
\mbq_\s=\mbj_{e\s}-\rho_\s\mbv_\s\left({mv_\s^2\over2}+w_\s\right )\,.
\label{heat}
\een
This definition subtracts the spin $\s$ energy current, the second term in the right-hand side of Eq.~\eqref{heat}, which contains the enthalpy per particle $w_\s$, related to the energy per particle $\e_\s$=$e_\s/\rho_\s$ by $\w_\s$=$\e_\s+p_\s/\rho_\s$.   Thus, the energy flux through the surface of a fluid element includes the work done by pressure forces $\oint p_\s\mbv_\s\cdot d\mbf{S}$, $d\mbf{S}$ being the normal vector surface,  which should be subtracted to obtain the heat current.\cite{landauFM}$^{,}$\cite{Note1}
It is also readily verified that this is the energy flux defined in Eq.~\eqref{densities} for a rigid shift $f_\mbps$$\to$$f_{\mbp-m\mbv_\s,\s}$ of the local Bose/Fermi distribution.
\begin{figure}[t]
\includegraphics[width=\linewidth]{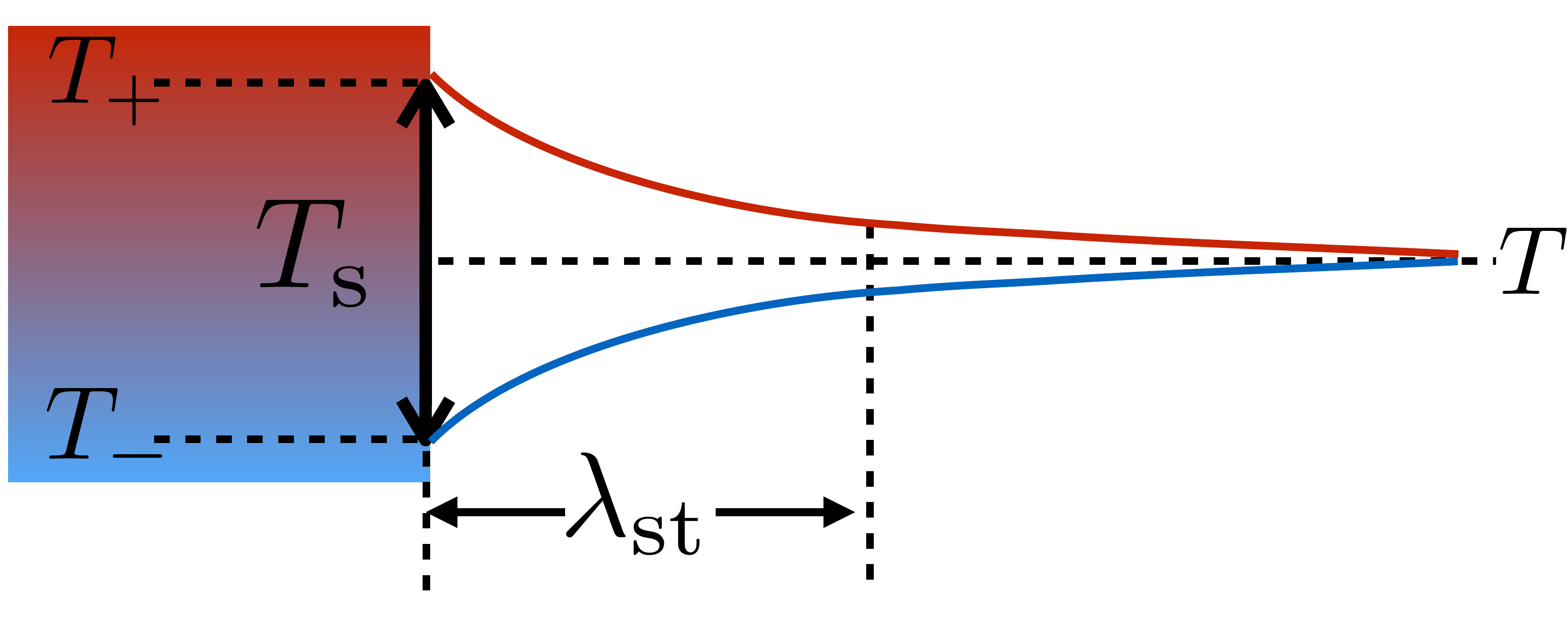}
\caption{(Color online) Illustration of spin-dependent temperature gradients and that decay on a length scale characterized by the spin-heat relaxation length $\lm_{st}$.}
\label{Tsfig}
\end{figure}

\section{spin-heat relaxation length and time\label{Ts}}

In this section, we use semi-phenomenological arguments to deduce the form of the spin-dependent temperature diffusion equations, which will define the spin-heat relaxation length and time, $\lm_{\rm st}$ and $\tau_{\rm st}$, respectively.  
We then report our results on the temperature and interaction strength dependence of these coefficients based on the solution of the Boltzmann equations Eq.~\eqref{boltz}.  Microscopic expressions for these coefficients are given in Sec.~\ref{spintemp}.

We first transform Eqs.~\eqref{entropy} into temperature diffusion equations, again following standard fluid mechanics,\cite{landauFM} but keeping track of the heat exchanges between spins.  We express the left-hand side of Eqs.~\eqref{entropy}, which represents the heat gained by spin $\s$ particles in a fluid element per unit volume per unit time, in terms of temperature derivatives as $\rho_\s c^\s_p (\p_t$$+$$\mbv_\s$$\cdot$$\bsdel) T_\s$,  where  $c^\s_p$=$T_\s({\p s}/{\p T_\s})_p$ is the heat capacity per particle at constant pressure, and assume linear-response heat currents $\mbq_\s$=$-\sum_{\tau=\pm}\kap'_{\s\tau}(T)\bsdel T_\tau$, where $\kap'_{\s\tau}(T)$ are the spin-dependent heat conductivities. Then, for the case of zero external forces ($\mbf{f}_\s$=$0$),  equal densities $\rho_+$=$\rho_-$=$\rho$, and zero total   particle current $\mbv_++\mbv_-=0$, Eq.~\eqref{entropy} becomes
\ben
\rho c_p \p_t T_\s=\sum_{\tau=\pm}\bsdel\cdot(\kappa'_{\s\tau}\bsdel T_{\tau})+\s\G_s
\label{Tsdiff}
\een
where in the left hand side we kept terms to leading order in the spin-heat accumulation $T_s$=$T_+-T_-$, and we defined $\G_s$=$\G_+-\G_-$.
Since $\G_s$ is a relaxation term for $T_s$, $\G_s$=0 when $T_s$=0, so that in linear response, it can be expanded as  $\G_s$=$-\rho c_pT_s/\tau_{\rm st}$, with $\tau_{\rm st}$ being the spin-heat relaxation time.\cite{Note2}
Taking the difference of the spin up and spin down components of Eq.~\eqref{Tsdiff} and specializing to the case of opposite temperature gradients, $\del T_+$=$-\del T_-$,  the  spin-heat diffusion equation reads
\ben
\p_tT_s={\kap_s'\over\rho c_p}\bsdel^2T_s-{T_s\over\tau_{\rm st}}\,.
\label{relax}
\een
where $\kap_s$=$\kap_{++}-\kap_{+-}$ is the spin-heat conductivity.\cite{Note3}
 In a steady state, the spin-heat diffusion length that sets the length on which $T_s\neq0$ is given by $\lm_s$=$\sqrt{\kappa_s'\tau_s/\rho c_p}$.
Such a steady state could be accomplished in practice, for example, by a laser spin-selectively heating  one side of the atomic cloud.   Such a situation approximated by the boundary condition of a fixed $T_s$ on the left side, with $T_s$ penetrating to a depth of $\lm_{\rm st}$ is  illustrated in Fig.~\ref{Tsfig}.  

\begin{figure*}[t]
\begin{center}
\includegraphics[width=0.7\linewidth]{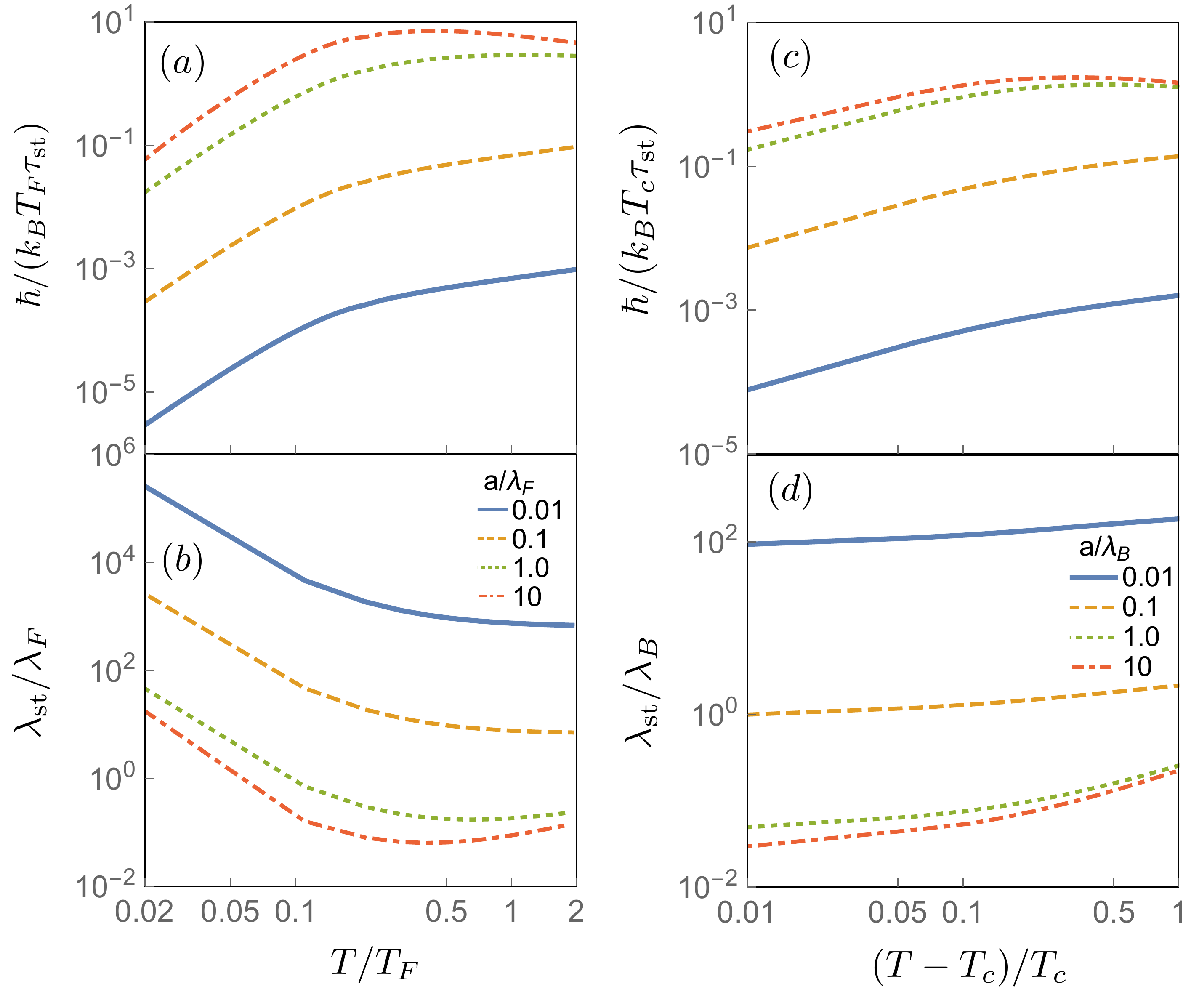}
\caption{(Color online) Left panel: Log-log plots of the normalized spin-heat relaxation rate and length for fermions, (a) $\hbar/k_BT_F\tau_{\rm st}$ and (b) $\lm_{\rm st}/\lm_F$, respectively, for the interspin scattering lengths $a/\lm_F$=$(0.01,0.1,1,10)$.
 Right panel: Log-log plots of the normalized spin-heat relaxation rate and length for bosons, (a) $\hbar/k_BT_c\tau_{\rm st}$ and (b) $\lm_{\rm st}/\lm_B$, respectively, for the interspin scattering lengths $a/\lm_B$=$(0.01,0.1,1,10)$.}
\label{STrelax}
\end{center}
\end{figure*}

In Fig.~\ref{STrelax}, we plot the normalized spin-heat relaxations rates and lengths, $\hbar/k_BT_c\tau_{\rm st}$ and $\lm_{\rm st}/\lm_B$ as a function of $(T-T_c)/T_c$ for bosons; $\hbar/k_BT_F\tau_{\rm st}$ and $\lm_{\rm st}/\lm_F$ as a function of $T/T_F$ for fermions, where $T_c$ is the temperature of Bose-Einstein condensation and $T_F$ is the Fermi temperature.  Here, we define  $\lm_B$=$\rho^{-1/3}$ for bosons, $\lm_F$=$2\pi(6\pi^2\rho)^{-1/3}$ is the Fermi wavelength for fermions, and $\rho$ is the equilibrium density.     The power-law dependence of the spin-heat relaxation coefficients on $T-T_c$ and $T/T_F$  is evident in the logarithmic plots.   These plots also suggest power-law behavior as a function of interspin scattering lengths for $a/\lm_B$$\leq$1 and $a/\lm_F$$\leq$1.  

For fermions, the spin-heat relaxation time ($\tau_{st}$) and length ($\lm_{st}$) diverges as $T/T_F$$\to$$0$.  This behavior is expected from Pauli blocking, the suppression of scattering due to Fermi statistics. 
Fig.~\ref{STrelax}~(b) shows relaxation lengths $\lm_{st}$ up to $10^5\lm_F$.  For a typical density of $\rho$=$10^{12}$ cm$^{-3}$, we have $ \lm_F$$\sim$$\lm_B$$\sim$1$~\mu$m. which gives $\lm_{\rm st}$$\sim$10$^{-2}$-1~mm for $a/\lm_F$$\sim$10-10$^3$, well within experimental resolution and much longer than spin-heat relaxation lengths found in the solid state.{\cite{dejeneNATP13}}$^{,}$\cite{Note4}
We note that at $T$=2$T_F$, one enters the high temperature limit of a classical two-component gas, and from Fig.~\ref{STrelax}(b), one finds $\lm_{\rm st}$$\sim$1$~\mu$m in this regime.

For bosons, the relaxation time also exhibits a power-law divergence as ${T\to T_c}$, as shown Fig.~\ref{STrelax}~(c), while the relaxation length $\lm_{\rm st}$ remains finite.  Fig.~\ref{STrelax}~(d)  shows that at ${(T-T_c)/T_c}$$\simeq$0.1 and $\lm_B$=1$~\mu$m, we have  $\lm_{\rm st}$$\sim$1$-$10$^2~\mu$m for $a/\lm_B$$\sim$10$^{-1}$$-$$10^{-2}$ and $\lm_{\rm st}$$\sim$10$^{-1}$$-$$10^{-2}~\mu$m for $a/\lm_B$$\sim$1$-$10.  Thus, for weak scattering $a/\lm_B$\textless1, $\lm_{\rm st}$ is within experimental resolution.\cite{Note5}

The physical interpretation for the behavior of the bosonic relaxation coefficients are less obvious.  At degenerate temperatures, one expects bosonic enhancement of scattering to be important.  This effect, for example, causes the spin-drag relaxation time to vanish as $T$$\to$$T_c$.\cite{duinePRL09,{drielPRL10}}  However, mathematically, the divergence in the heat capacity $c_p$ as $T$$\to$$T_c$ dominates over the Bose enhancement of the relaxation integral $\G_s$, causing the relaxation length $\tau_{\rm st}$ to diverge.  Physically, the diverging heat capacity indicates that an increasing amount of heat is necessary per temperature change as $T$$\to$$T_c$, which stabilizes the spin-heat accumulation.
On the other hand, the relaxation length $\lm_{\rm st}$$\sim$$\sqrt{\kap'_s/\G_s}$  decreases as $T$$\to$$T_c$ on account of Bose enhancement, but for weak scattering this enhancement is weak enough so that $\lm_{\rm st}$ remains finite even near $T_c$.

 In the remainder of the paper, we present a detailed computation of the spin-heat relaxation coefficients and the related spin-heat transport coefficients.  
  
\section{Linearized Boltzmann equations \label{linear}}
In this section, we will use the linearized Boltzmann equation to derive microscopic expressions for spin-dependent particle and heat transport and relaxation coefficients, in linear response to perturbations from equilibrium, i.e., gradients in temperature and density, and external forces.  We will outline a general method for obtaining the linear-response coefficients for generic spin-dependent forces, and clarify the relation between our work and that of Ref.~[\onlinecite{kimPRA12}] and the corresponding classical problem in the literature [\onlinecite{chaikinBook00}].  We then specialize to the response to spin-forces and spin-heat accumulation gradients, which is the focus of this work.

We will employ the standard Chapman-Enskog expansion of the non-equilibrium distribution,\cite{chapmanMTNG70}
\begin{align}
 n_ {\mbp\s}(\mbr,t)&=f_ {\mbp\s}(\mbr,t)-\p_\e f^0_\mbps \Phi_{\mbps}(\mbr)\,;\nn
f^0_\mbps&={1\over \exp[(\e_\mbp-\mu_\s)/k_BT]-\zeta}\,, \nn
\p_\e f^0_\mbps&=-{f^0_\mbps(1+\zeta f^0_\mbps)\over k_BT}\,,\nn
 f_ {\mbps}(\mbr,t)&={1\over \exp[(\e_{\mbp}-\mu_\s(\mbr,t))/ k_BT_\s(\mbr,t)]-\zeta}\,,
  \label{ansatz}
\end{align}
where $\e_\mbp=\mbp^2/2m$ is the free-particle dispersion, $f^0_\mbps$ and $f_\mbps(\mbr,t)$ are the global and local equilibrium Bose(Fermi) distribution, respectively, $\mu_\s(\mbr,t)$ and $T_\s(\mbr,t)$ are the local, spin-dependent chemical potentials and temperatures which are determined by the local particle and energy densities, cf.~Appendix \ref{thermo}, while the perturbed distribution $\Phi_\mbps(\mbr)$ describes the response to spatial inhomogeneities. The  parametrization of the distribution function in Eq.~\eqref{ansatz} represents the leading order expansion in the ratio of the mean free path to spatial gradients, called the Knudsen number,\cite{landauPK} which is assumed to be small in our perturbative solution, to be given in following sections.   


\subsection{Response to generic spin-dependent forces}
Substituting the expansion Eq.~\eqref{ansatz} in the Eq.~\eqref{boltz}, the Boltzmann equation separates into an equation of order $\bsdel$ and one of $(\p_t,\bsdel^2)$.  To leading order in $T_s$, they are
\begin{align}
&{w_\s-\e_\mbp\over T}(\mbv_\mbp\cdot\bsdel) T_\s+\mbv_p\cdot\mbf{F}_\s=\frac{\mcal{C}_{\mbps}[\vec{\Phi}(\mbr)]}{\p_\e f^0_\mbps}\,,\label{boltz1a0}\\
&{w_\s-\e_\mbp\over T}\p_t T_\s+(\mbv_\mbp\cdot\bsdel+\mbf{f}_\s\cdot\del_\mbp)\Phi_\mbps(\mbr)=
\frac{\mcal{C}_{\mbp\s}[\vec{f}(\mbr,t)]}{\p_\e f^0_\mbps}\,,
\label{boltz1b}
\end{align}
where $w_\s$ is the enthalpy per particle, we introduced the spin $\s$ thermodynamic force $\mbf{F}_\s$$=$$\mbf{f}_\s-\bsdel p_\s/\rho_\s$.  Following standard conventions,\cite{landauPK} in the advective terms in the left hand side of Eq.~\eqref{boltz1a0},
we choose pressure and temperature as our independent variables, with changes in the chemical potential $\mu_\s$=$\mu_\s(p_\s,T_\s)$ given by the Gibbs-Duhem relation,
\[d\mu_\s=-s_\s dT_\s+dp_\s/\rho_\s\,,\]
 where $s_\s$ and $p_\s$ are the spin $\s$ entropy per particle and pressure, respectively.  We then eliminated $\mu_\s$ in favor of $w_\s$ using the thermodynamic identity $w_\s$$=$$\mu_\s+Ts_\s$.  
\app{thermo}summarizes some thermodynamic properties of the equilibrium gas. 

Linearizing the collision integral in Eq.~\eqref{boltz1a0} with respect to $\Phi_\mbps$ and performing the integrations over final momenta $\mbp_3$ and $\mbp_4$, we have
\begin{align}
\mcal{C}_{\mbp_1\s}[\vec{\Phi}]&=- f^0_{\mbp_1\s}\int{d\mbp_2\over(2\pi\hbar)^3}|\mbv_r|\nn
&\quad \int d\Omega_r'\sum_{\tau=\pm}{d\s_{\s\tau}\over d\Omega'_r}f^0_{\mbp_2\tau}  (1+\zeta f^0_{\mbp_3\s})(1+\zeta f^0_{\mbp_4\tau})\nn
&\quad(\Phi_{\mbp_3\s}+\Phi_{\mbp_4\tau}-\Phi_{\mbp_1\s}-\Phi_{\mbp_2\tau}),
\label{lincoll}
\end{align}
where $\mbv_r$=$2\mbp_r/m$ is the relative velocity,\cite{Note6}
$\W'_r$ are the spherical angles of $\hatbp_r'$.   
In the integrand of Eq.~\eqref{lincoll}, energy and momentum conservation has been enforced, so that $\mbp_{3}$=$\mbP/2+\mbp_r'$, $\mbp_{4}$=$\mbP/2-\mbp_r'$, where $\mbP$=$\mbp_1+\mbp_2$ is the center-of-mass momentum and $|\mbp_r|$=$|\mbp_r'|$ is the relative momentum. We note that this linearized collisional possesses the same collisional invariants given in Eq.~\eqref{Cinv}.

Following the standard approach for solving the Boltzmann equation,\cite{reichlSP98} we first solve Eq.~\eqref{boltz1a0} and \eqref{lincoll} for the perturbed distributions $\Phi_{\mbps}$, and then substitute them into Eq.~\eqref{boltz1b}.  At the level of the hydrodynamic equations, this procedure corresponds to substituting the linear response currents due to $\Phi_{\mbps}$  into the continuity equations, Eq.~\eqref{particle}, \eqref{euler}, and \eqref{energy}, resulting in diffusion equations that govern the spatial-temporal dependence of $T_\s$ and $\mbF_\s$.   The collision integral in Eq.~\eqref{boltz1b} leads to the spin-heat relaxation term $\G_\s$, and is nonzero only for the nonequilibrium part $f_{\mbps}(t)$ due to the spin-heat accumulation $T_s$.  In Sec.~\ref{spintemp}, we will expand this term to leading order in $T_s$ to compute the relaxation time.
We note that Eq.~\eqref{boltz1a0} does not contribute to the energy continuity equation since  $\int d\mbp\,\e_\mbp \times$Eq.~\eqref{boltz1a}=0 because the left hand side vanishes by isotropy and $\Gamma_\s[\p_\e f\Phi_\mbps]=0$.

To solve Eq.~\eqref{boltz1a0}, we parametrize the perturbed distribution in linear response as 
\begin{align}
\Phi_{\mbps}
&=\sum_{\tau}\bm{\Phi}_F^{\st}(\mbp) \cdot\mbf{F}_{\tau}+\bm{\Phi}_T^{\st}(\mbp)\cdot(-k_B\bsdel T_{\tau})\,,\nn
&\equiv\sum_{\al,\tau}\bm{\Phi}_\al^{\st}\cdot\mbf{X}_{\al\tau}\,.
\label{phis}
\end{align}
where $\al$=$F,T$ labels the thermodynamic forces $\mbf{X}_{F\s}$=$\mbF_\s$ and  $\mbf{X}_{T\s}$=$-k_B\bsdel T_\s$.  By symmetry, we have $\bm{\Phi}_\al^{+-}$=$\bm{\Phi}_\al^{-+}$.
Substituting Eq.~\eqref{phis} into the linearized collision integral Eq.~\eqref{lincoll}, we have
\begin{align}
\mcal{C}_{\mbps}[\vphi]&=
\sum_{\al,\tau}\bm{\mcal{C}}_{\mbp\al}^{\s\tau}\cdot\mbf{X}_{\al\tau}\,,\,\,
\bm{\mcal{C}}_{\mbp\al}^{\s\tau}\equiv \mcal{C}_{\mbps}[\bm{\Phi}_\al^{+\tau},\bm{\Phi}_\al^{-\tau}]\,.
\label{lincolla}
\end{align}
Noting that  $\mbf{F}_\s$ and $\bsdel T_\s$ are linearly independent,  Eq.~\eqref{boltz1a0} separates into two equations
\begin{align}
\mbv_\mbp\cdot \mbF_\s&=\sum_{\tau}\frac{\bm{\mcal{C}}_{\mbp F}^{\s\tau}}{\p_\e f^0_\mbps }
\cdot\mbF_{\tau}\,,\nn
\left({\e_\mbp-w_\s\over k_BT}\right)
 \mbv_\mbp\cdot\bsdel T_\s&=\sum_{\tau}\frac{\bm{\mcal{C}}_{\mbp T}^{\s\tau}}{\p_\e f^0_\mbps }  \cdot\bsdel T_{\tau}\,.
\label{boltz1a}
\end{align}

Before solving Eq.~\eqref{boltz1a} for $\Phi_\mbps$ using the method described in Sec.~\ref{moment}, we consider the structure of the response coefficents that follow from them.  The particle and heat currents, $\mbj_\s$ and $\mbq_\s$, respectively, are given by
\begin{align}
\left(\begin{array}{c}{\mbj}_\s\\{\mbq}_\s\end{array}\right)
&\equiv\left\langle\left(\begin{array}{c}1\\\e_\mbp-w_\s\end{array}\right)
\mbv_\mbp\Phi_{\mbp \s}\right\rangle\nn
&=\sum_{\tau}
\left(\begin{array}{cc}
 {\mcal{L}}^{\s\tau}_{FF}  & \mcal{L}^{\s\tau}_{FT}\\
  k_BT\mcal{L}^{\s\tau}_{TF}  & k_BT\mcal{L}^{\s\tau}_{TT} 
\end{array}\right)
\left(\begin{array}{c}
 \mbf{F}_\tau  \\
- k_B\bsdel T_\tau
\end{array}\right)\nn
&=\sum_{\be\tau}\coltwo{\mcal{L}^{\s\tau}_{F\be}}{k_BT\mcal{L}^{\s\tau}_{T\be}}\mbf{X}_{\be\tau}\,.
\end{align}
where we defined a set of response coefficients ${\mcal{L}}^{\s\tau}_{\al\beta}$ with appropriate factors of $T$ are taken out for convenience.  Defining the momentum-space inner product
\begin{align}
\langle\Phi_\s\chi_\s\rangle&\equiv-\int{d^3p\over(2\pi\hbar)^3}\p_\e f^0_\mbps\,\,\Phi_\mbps\chi_\mbps\,,
\end{align} 
for a generic spin and momentum dependent function $\chi_\mbps$, the response coefficients are given by
\[\mcal{L}^{\s\tau}_{F\be}=\langle\mbv_\mbp\otimes\bm{\Phi}^{\s\tau}_\be\rangle\,,\quad
\mcal{L}^{\s\tau}_{T\be}=\left<{\e-w_\s\over k_BT}\mbv_\mbp\otimes\bm{\Phi}^{\s\tau}_\be\right> \,,\]
where $\otimes$ denotes the vector tensor product, though ${\mcal{L}}_{\al\beta}$ will be diagonal in real space since we do not consider any Hall effects in this paper. We note that $\mcal{L}_{\al\be}^{+-}$=$\mcal{L}_{\al\be}^{-+}$ in the cases we study here (without time-reversal symmetry breaking), so we only have three independent spin-resolved coefficients.  Onsager reciprocity, which we will prove below, also 
requires these coefficients to be symmetric in the space of thermodynamic forces, hence $\mcal{L}_{FT}^{\st}$=$\mcal{L}_{TF}^{\st}$.

Since we are interested in the spin response, we next transform the response matrix to the total particle $(t)$ and spin $(s)$ sectors.  Although in general, the matrix of linear response coefficients have couplings in the $4\times4$ space of spins and thermodynamic forces, we will consider the case in which the spin and total particle response decouple.   Consider the response of the total ($\mbj$) and spin currents ($\mbj_s$) (in units of $\hbar/2$) in response to average ($\mbF$) and spin forces ($\mbF_s$) defined by $\mbF_\s$=$\mbF+\s\mbF_s/2$,  given by
\begin{align}
\left(\begin{array}{c}\mbj_t \\\mbj_s \\ \end{array} \right)&=
\left( \begin{array}{c}
 \mbj_++ \mbj_- \\
 \mbj_+- \mbj_- 
\end{array}
\right)=
\left(
\begin{array}{cc}
\mcal{L}^{(t)}_{FF}&\mcal{L}^{(st)}_{FF}/2\\
\mcal{L}^{(st)}_{FF}&{\mcal{L}}^{(s)}_{FF}
\end{array}
\right)\coltwo{\mbf{F}}{\mbF_s}\,;\nn
\mcal{L}^{(t)}_{FF}&=\mcal{L}_{FF}^{++}+\mcal{L}_{FF}^{--}+2\mcal{L}_{FF}^{+-}\,,\nn
\mcal{L}^{(ts)}_{FF}&=\mcal{L}_{FF}^{++}-\mcal{L}_{FF}^{--}\,,\nn
{\mcal{L}}^{(s)}_{FF}&=\frac{\mcal{L}_{FF}^{++}+\mcal{L}_{FF}^{--}}{2}-\mcal{L}_{FF}^{+-}\,,
\label{response}
\end{align}
and the analogous response matrix holds for the temperature gradient response $\mcal{L}^{\s\tau}_{TT}$ and cross response $\mcal{L}^{\s\tau}_{FT}$. We note that our spin current in general differs from the relative current sometimes defined in the literature of two-component gases, \cite{chaikinBook00}
\[\mbj_{\rm rel }=\mbj_s-(\rho_s/\rho)\mbj\,,\quad \rho_s=\rho_+-\rho_- \, \]
 which subtracts the spin current carried by the average velocity of the fluid when it is spin-polarized,  though in the unpolarized case studied below they are equal.  {Our definition of the spin current has the advantage that it is the current that couples to spin-dependent potentials in the Hamiltonian, which makes it convenient for comparison with calculations using the Kubo formula.}  

The spin and total particle response decouples when the intraspin response coefficients are equal $\mcal{L}_{\al\be}^{++}$=$\mcal{L}_{\al\be}^{--}$, so that $\mcal{L}_{\al\be}^{(st)}$=$0$,  which means that the center-of-mass motion of the atomic cloud is decoupled from the relative motion of its components, i.e., the spin currents.  In this case, the remaining two independent spin-resolved coefficients, given by $\mcal{L}^{(t)}_{\al\be}$=$2(\mcal{L}_{\al\be}^{++}+\mcal{L}_{\al\be}^{+-})$ and  $\mcal{L}^{(s)}_{\al\be}$=$\mcal{L}_{\al\be}^{++}-\mcal{L}_{\al\be}^{+-}$, can be  determined by the response for  two cases: (i) when the average components are zero, $\del T_+$=$-\del T_-$ and $\mbF_+$=$-\mbF_-$, and (ii) when the spin components are zero, $\del T_+$=$\del T_-$ and  $\mbF_+$=$\mbF_-$.  The latter was recently studied in the high-temperature limit in Ref.~[\onlinecite{kimPRA12}], where the ``spin-Seebeck" coefficient is proportional to $\mcal{L}^{(st)}_{FT}$. In the next section, we will consider the former case of opposite forces,  which drives \emph{pure} spin currents stabilized by interspin scattering.  

\subsection{Response in an unpolarized gas\label{unpolarized}}

Henceforth, we consider the  case of equal equilibrium densities (and masses), $\rho_+$=$\rho_-$$\equiv$$\rho$, and a local-equilibrium distribution with a spin-heat accumulation gradients $T_s(\mbr,t)$, as shown in Fig.~\ref{Tsfig}.    Furthermore, we consider equal intraspin scattering cross sections ${d\s_{++}/ d\Omega}$=${d\s_{--}/ d\Omega}$.  Then, by symmetry, $\mcal{L}_{\al\be}^{++}$=$\mcal{L}_{\al\be}^{--}$, so that the off-diagonal coupling in Eq.~\eqref{response} vanishes, $\mcal{L}_{\al\be}^{(st)}$=$0$.  The linearized Boltzmann equation for the total and spin distributions, $n_\mbp$ and $n_{\mbp s}$, respectively defined by $n_\s$=$(n+\s n_s)/2$, also decouples.  Defining the corresponding average and spin components 
\ben
\Phi_\mbps={\phi_{\mbp t}+\s \phi_{\mbp s}\over2}\,,\quad \mcal{C}_\mbps={C_{\mbp t}+\s C_{\mbp s}\over2}\,,
\een
the collision integrals are given by
\begin{align}
&\left(\begin{array}{c}
C_{\mbp_1 t}\\C_{\mbp_1 s}
\end{array}\right)
=-\int{d\mbp_2\over(2\pi\hbar)^3}|\mbv_r|f^0_1f^0_2\\
&\qquad\qquad\qquad\int d\Omega_r' (1+\zeta f^0_3)(1+\zeta f^0_4)\nn
&\left(\begin{array}{c}
(d\s_{++}/d\Omega'_r+d\s_{+-}/d\Omega'_r)\D_{++}[\phi_{\mbp t}]\\
(d\s_{++}/ d\Omega'_r)\D_{++}[\phi_{\mbp s}]+(d\s_{+-}/ d\Omega'_r)\D_{+-}[\phi_{\mbp s}]
\end{array}\right)\nonumber\,,
\label{lincoll1}
\end{align}
where  we introduced the notation 
\begin{align}
\D_{++}[\chi_\mbp] &=\chi_{\mbp_3}+\chi_{\mbp_4}-\chi_{\mbp_1}-\chi_{\mbp_2}\nn
\D_{+-}[\chi_\mbp]&=\chi_{\mbp_3}-\chi_{\mbp_4}-\chi_{\mbp_1}+\chi_{\mbp_2}\,,
\end{align}
for a generic momentum dependent function $\chi_\mbp$.
The intraspin term satisfies $\D_{++}[\mbp]=0$ and $\D_{++}[\e_\mbp]=0$, reflecting  momentum and energy conservation, while $\D_{++}[1]$=$0$ and $\D_{+-}[1]$=$0$. 
 We note that from Eq.~\eqref{lincoll1}, the Boltzmann equation for the total distribution is the same as that of a one-component gas with the interspin and intraspin differential cross sections added together, and the corresponding problem has been studied extensively in the literature.\cite{landauPK}  Henceforth, we focus on the spin component.


The spin components of Eqs.~\eqref{phis} and \eqref{lincolla} are
\begin{align}
\phi_{\mbp s}&=\bs{\phi}_{F}(\mbp) \cdot\mbf{F}_s+\bs{\phi}_{T}(\mbp)\cdot(-k_B{\bsdel T_s })\,,\nn
C_{\mbp s}[\phi_s]&= C_{\mbp s}[\bs{\phi}_{F}] \cdot\mbf{F}_s+ C_{\mbp s}[\bs{\phi}_{T}] \cdot(-k_B\bsdel T_s)\,,
\end{align} 
and the spin component of Eq.~\eqref{boltz1a} is
\ben
\mbv_\mbp=\frac{C_{\mbp s}[\bs{\phi}_{F}]}{\p_\e f^0_\mbp }\,,\quad 
\left({\e_\mbp-w\over k_BT}\right)\mbv_\mbp=\frac{C_{\mbp s}[\bs{\phi}_{T}]}{{\p_\e f^0_\mbp }}\,,
\label{boltz1}
\een
where we write $\phi_{\mbp s}$$\equiv$$\sum_\al\bs{\phi}_\al\cdot\mbf{X}^{(s)}_\al$, $\mbf{X}^{(s)}_\al$ being the spin component of the thermodynamic forces.  
Since the collision integral is a linear integral operator, it will be convenient to introducing the notation 
\ben
\hat{C}_{ s}\chi_\mbp\equiv\frac{{C}_{\mbp s}[\chi]}{ \p_\e f^0_\mbp}\,,
\label{Csop}
\een
for a generic momentum dependent function $\chi_\mbp$.
then Eq.~\eqref{boltz1} can regarded as an eigenvalue equation for the collision integral operator, and solving it amounts to inverting the collision operator $\hatC_s$.  
In Sec.~\ref{moment}, we will solve Eq.~\eqref{boltz1a} for $\bs{\phi}_{F}$ and $\bs{\phi}_{T}$  using a moment expansion. 

The spin and spin-heat currents are given by
\begin{align}
\left(\begin{array}{c}{\mbj}_s\\{\mbq}_s\end{array}\right)
&\equiv\left\langle\left(\begin{array}{c}1\\\e_\mbp-w\end{array}\right)
\mbv_\mbp\phi_{\mbp s}\right\rangle\nn
&\equiv
\left(\begin{array}{cc}
 {\mcal{L}}^{(s)}_{FF}  & k_B\mcal{L}^{(s)}_{FT}\\
  k_BT\mcal{L}^{(s)}_{TF}  & k_B^2T\mcal{L}^{(s)}_{TT} 
\end{array}\right)
\left(\begin{array}{c}
 \mbf{F}_s  \\
- \bsdel T_s
\end{array}\right)\,,\nn
\langle\phi\chi\rangle&\equiv-\int{d^3p\over(2\pi\hbar)^3}\p_\e f^0_\mbp\,\,\phi_\mbp\chi_\mbp\,.
\label{currents}
\end{align}
Henceforth, we drop the superscript $(s)$ for $L_{\al\be}=\mcal{L}^{(s)}_{\al\be}$.  The response coefficients are given in terms of the spin distributions by
\begin{align}
L_{F\be}&=\langle\mbv_\mbp\otimes\bs{\phi}_\be\rangle=\<\hatC_s\bs{\phi}_F\otimes\bs{\phi}_\be\>\,,\nn
L_{T\be}&=\left<{\e_\mbp-w\over k_BT}\mbv_\mbp\otimes\bs{\phi}_\be\right>=\<\hatC_s\bs{\phi}_T\otimes\bs{\phi}_\be\> \,.
\label{L}
\end{align}
In the Eq.~\eqref{L},  we expressed $L_{\al\beta}$ in terms of the collision integrals using Eq.~\eqref{boltz1}.  They are symmetric by the symmetry of the collision integral operator,\cite{landauPK}
\[L_{FT}=\<\hatC_s\bs{\phi}_{F}\otimes\bs{\phi}_{T}\>=\<\bs{\phi}_{F}\otimes\hatC_s\bs{\phi}_{T}\> =L_{TF}\,,\]
and thus satisfy the Onsager reciprocity principle.

Finally, it is conventional to define the transport coefficients by
\begin{align}
\left(\begin{array}{c}
 \mbj_s \\  \mbq_s
\end{array} \right)
&\equiv\s_s
\left( \begin{array}{cc}
1&  S_s\\
 P_s  & \kappa_s' /\s_s
\end{array}\right)
\left(\begin{array}{c}
 \mbf{F}_s  \\
- \bsdel T_s
\end{array}\right)\,,
\label{currents1}
\end{align}
which are related to the coefficients $L_{\al\be}$ by
\ben
\s_s=L_{FF}\,,\,\, S_s={P_s\over T}=k_B{L_{FT}\over L_{FF}}\,,\,\,\kappa_s'=k_B^2TL_{TT}\,.
\een
Furthermore, we define the spin-heat conductivity at zero spin current $\kappa_s$ and a figure of merit for thermo-spin conversion $Z_{\rm s} T$  given by\cite{mahanSSP97}
\begin{align}
\kappa_s&=\kappa_s'-\s_sS_s^2T=k_B^2T{\det \hat{L}\over L_{FF}}\,,\nn
Z_sT&=\frac{\s_sS_s^2T}{\kappa_s}=\frac{L^2_{FT}}{\det\hat{L}}\,,
\label{kap}
\end{align}
where $\hat{L}$ is the matrix of response coefficients $L_{\al\be}$.
The response coefficients in \eqref{currents1}  can be accessed directly in experiments.  The spin-Seebeck effect for example, can be measured in the manner discussed in Ref.~[\onlinecite{wongPRA12}]. 


\subsubsection{Relaxation coefficients\label{spintemp}}
We now derive a microscopic expression for the spin-heat relaxation rate  $1/\tau_{\rm st}$ and length $\lm_{\rm st}$.  As mentioned previously, this relaxation term comes from the energy transfer between spins represented by $\Gamma_\s[\de \vec{f}_{\mbp}]$ in Eq.~\eqref{Gamma}, where $\de \vec{f}_{\mbp}$ is the perturbation to the local equilibrium distribution due to the spin-heat accumulation $T_s$, given to leading order by 
\[\de f_{\mbps}(t)={\s T_s\over2}\p_\e f_\mbps^0{\e_\mbp-w\over k_B T}\,.\]
Recalling that the energy is a collisional invariant cf.~[Eq.~\eqref{Cinv}], in $\mcal{C}_{\mbps}[\de f_{\mbps}(t)]$, only the spin component $C_{\mbp s}$ is nonzero, thus 
\[\G_\s[\de \vec{f}_\mbp]={\s T_s\over2}\<\e_\mbp\,\hatC_s(\e_\mbp/k_BT)\>\,.\] 
Thus, the spin-heat relaxation rate and length, cf.~Eq.~\eqref{relax}, is given by 
\begin{align}
{1\over\tau_{\rm st}}&=\frac{\<\e_\mbp\,\hatC_s(\e_\mbp/k_BT)\>}{\rho c_p}\,,\nn
{\lm_{\rm st}}&=\sqrt{\frac{\kappa_s'\tau_{\rm st}}{\rho c_p}}
=\sqrt{\frac{k_B^2T{L}_{TT}}{\<\e_\mbp\,\hatC_s(\e_\mbp/k_BT)\>}}\,.
\label{Eq:STrelax1}
\end{align}
\subsubsection{Total entropy production \label{totentropy}}

In this section, we compute the total entropy production due to spin-heat accumulation gradients and spin forces. 
It can be conveniently computed directly from the non-equilibrium entropy density defined in terms of the distribution function, given by
\[\rho_\s s_\s= {k_B}\intdppp  [\z(1+\zeta n_\mbps)\ln(1+\zeta n_\mbps)-n_\mbps \ln n_\mbps]\,.\]
The entropy density production is thus
\[\p_t(\rho_\s s_\s)={k_B}\intdppp\p_t{n}_\mbps\ln\left({1+\zeta n_\mbps\over n_\mbps}\right)\,.\]
The total entropy production is the equation above integrated over all space.  It contains contributions only from the collision integral. Using the Boltzmann equation Eq.~\eqref{boltz} and the ansatz Eq.~\eqref{ansatz}, we find the heating given by the quadratic form,
\begin{align}
&T\sum_\s\int d\mbr\, \p_t(\rho_\s s_\s)_{\rm coll}={1\over2}\int d\mbr\,\<\phi_{\mbp s}\hatC_s\phi_{\mbp s}\>\\
&={1\over2} \int d\mbr\,\sum_{\al,\beta} \mbf{X_\al}\cdot{L}_{\al\beta}\cdot\mbf{X}_\beta\nn
&=\int d\mbr\,\left({L_{FF}\over2}\mbf{F}_s^2+{k_B^2L_{TT}\over2}{\bsdel{T}_s}^2+k_BL_{FT}\mbf{F}_s\cdot{\bsdel T_s}\right)\,.\nonumber
\end{align}
Measurement of this heating will provide an indirect measurement of the response coefficients $L_{FF}$, $L_{FT}$, and $L_{TT}$.

\section{Solution by moment expansion\label{moment}}

To solve the steady-state Boltzman equation Eq.~\eqref{boltz1}, we use a polynomial expansion,\cite{Note7}  
\ben
\bs{\phi}_\al =\sum_{n=0}  c_n^{(\al)}\left({\e_\mbp\over k_BT}\right)^n\mbp\,,\quad \al=F,T\,.
\label{soln}
\een
Taking the $m$th moment by
\[-\int{d^3p\over(2\pi\hbar)^3} \pfrac{\e_\mbp }{k_BT}^n\mbp \cdot {\rm Eq. \eqref{boltz1}},\] with $m=0,1,2,\ldots$, results in a set of equations
\ben
{3\over\Lm^3}l^{(\al)}_m=\sum_{n=0}^\infty \mathsf{C}_{mn}c^{(\al)}_n\,,
\label{moments}
\een
 where the matrix elements of the collision operator are 
 \begin{align}
 \mathsf{C}_{mn}&=\left<\pfrac{\e_\mbp }{k_B T}^m\mbp\cdot\hat{C}_s\pfrac{\e_\mbp }{k_B T}^n\mbp\right>\label{cmat}\\
 &=-\int{d^3p\over(2\pi\hbar)^3}   \left({\e_\mbp \over k_B T}\right)^m \mbp\cdot C_{\mbp s}\left[\left({\e_\mbp \over k_B T}\right)^n\mbp\right]\,,\nonumber
\end{align}
where $\cdot$ denotes a dot product, and we defined the following functions 
\begin{align}
l_m^{(F)}&=l_m,\quad l_m^{(T)}=l_{m+1}-{w\over k_B T} l_m\,,\\
l_n&\equiv{\Lm^3}\left\langle\mbp\otimes\mbv_p\pfrac{\e_\mbp}{k_BT}^n\right\rangle=\zeta{\Gamma_{n+5/2}\over\Gamma_{5/2}}{\rm Li}_{n+3/2}(\zeta z)\,,\nonumber
\end{align} 
where $\Lambda$=$\hbar\sqrt{2\pi/mk_BT}$ is the deBroglie wavelength, $z$=$e^{\mu}/T$ is the fugacity, ${\rm Li}_s(z)$=$\sum_{n=1}^{\infty}z^n/n^s$ are the polylogarithmic functions, and $\G_n$ denotes the Gamma function. The first two $l_n$'s can be expressed in terms of thermodynamic quantities, cf.~appendix \ref{thermo},
\ben
\begin{pmatrix}l_0\\l_1\end{pmatrix}
=\zeta\begin{pmatrix}{\rm Li}_{3/2}(\zeta z)\\{5\over2}{\rm Li}_{5/2}(\zeta z)\end{pmatrix}
=\rho\Lm^3
\begin{pmatrix}1\\w/k_BT\end{pmatrix}\,.
\label{ln}
\een
The expansion coefficients that follows from inverting Eq.~\eqref{moments} are 
\ben
c_n^{(\al)}(z,\Lm)={3\over\Lm^3}\sum_{m} \mathsf{C}^{-1}_{nm}(z,\Lm)l^{(\al)}_m(z)\,.
\label{ab}
\een
The response coefficients follows from substituting the expansion Eq.~\eqref{soln} in Eq. \eqref{L},\cite{Note8}
\ben
L_{\al\be}={1\over\Lm^3}\sum_n c_n^{(\be)} l^{(\al)}_n\,.
\label{Lab}
\een
Truncating this expansion at the second order,\cite{Note9} the $L_{\al\beta}$ coefficients are
\begin{align}
L_{FF}&=\rho\left(c_0^{(F)}+c_1^{(F)}{w\over k_B T}\right)\,,\nn 
L_{FT}&=\rho\left(c_0^{(T)}+c_1^{(T)}{w\over k_B T}\right)\,,\nn
L_{TF}&=\rho c_1^{(F)} f(z,T)\,,\nn
L_{TT}&=\rho c_1^{(T)} f(z,T)\,,\nn
f(z,T)&\equiv\frac{l_1^{(T)}}{\rho\Lm^3}=\left[{35\over4}{ {\rm Li}_{7/2}(\zeta z)\over {\rm Li}_{3/2}(\zeta z)}-\pfrac{w}{k_B T}^2 \right]\,.
\label{Lab1}
\end{align}
Note that since $l_0^{(T)}$=$0$, the heat current, proportional to $L_{TT}$ and $L_{TF}$, does not depend on $c_0^{(F)}$ and $c_0^{(T)}$.  A comparison of this solution with the one used to compute the spin-drag relaxation time the literature in the absence of spin-heat currents is given in Appendix \ref{spindrag}.\cite{duinePRL09,drielPRL10} 

We conclude this section by verifying that our approximate solution satisfys Onsager reciprocity.   Using Eq.~\eqref{ab} to express the transport coefficients in terms of the collision matrix elements,
\begin{align}
L_{FF}&={3\over\Lm^6}\sum_{mn} \mathsf{C}^{-1}_{nm}l_nl_m\,,\label{L2}\\
 L_{FT}&={3\over\Lm^6}\sum_{mn} \mathsf{C}^{-1}_{nm}l_nl_{m+1}-{w\over k_B T}L_{FF} \nn
L_{TF}&={3\over\Lm^6}\sum_{mn} \mathsf{C}^{-1}_{nm}l_{n+1} l_m-{w\over k_B T}L_{FF}\,,\nn
 L_{TT}&={3\over\Lm^6}\sum_{nm} \mathsf{C}^{-1}_{nm}(l_{n+1}l_{m+1}-2{w\over k_B T}l_{n}l_{m+1})\nn
 &\qquad+\left({w\over k_B T}\right)^2L_{FF}\,.\nonumber
\label{L2}
\end{align}
Since $\mathsf{C}_{nm}$ is symmetric, so is  $\mathsf{C}^{-1}_{nm}$, and thus we  satisfy the Onsager relation, $L_{FT}=L_{TF}$, order by order in this expansion.

\section{Transport and relaxation coefficients for s-wave scattering\label{swave}}

\begin{figure*}[t]
\begin{center}
\includegraphics[width=.7\linewidth]{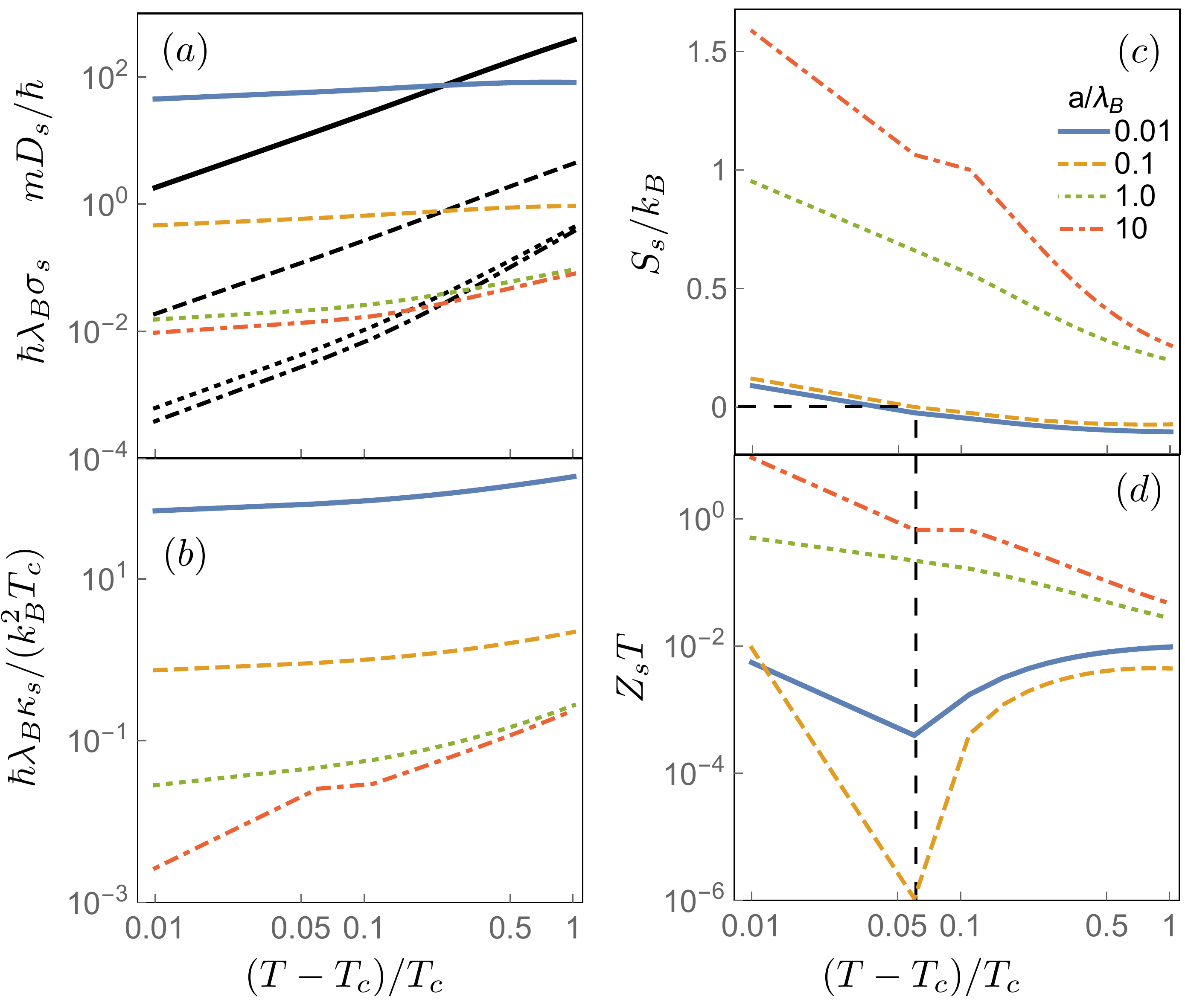}
\caption{(Color online) Bosonic spin-heat transport coefficients for the ratios of interspin scattering lengths to interparticle spacing $a/\lm_B$=$(0.01,0.1,1,10)$. (a) In color:  Log-log plot of the normalized spin conductivity $\hbar\lm_B\s_s$.  In black: Log-log plot of the normalized spin diffusitivty $\hbar D_s/m$ . (b) Log-log plot of the normalized spin-heat conductivity $\hbar\lm_B\kap_s/k_B^2T_c$. (c) Log-linear plot of the spin-Seebeck coefficient in units of the Boltzmann constant, $S_s/k_B$.  Dashed black line indicates the zero crossing for weak scattering lengths $a/\lm_B$=$(0.01,0.1)$.  (d) Log-log plot of the spin-heat figure of merit $Z_sT$.}  
\label{BoseCoefs}
\end{center}
\end{figure*}
In this section, we present our results for spin-heat transport coefficients as a function of temperature and interaction strength for the $s$-wave scattering differential cross section.  We numerically evaluate the second order formulas for the response coefficients given in Eq.~\eqref{Lab1}.  The computation of the required collision matrix elements $\mathsf{C}_{nm}$ is detailed in appendix \ref{collisionMat}.  


It will be helpful in understanding our numerical results to use scaling arguments to deduce the form of transport coefficients as a function of temperature and scattering length. We first factor out the generic temperature dependence by expressing the collision matrix elements as a function of the dimensionless momentum $\til{\mbp}$=$({\Lambda/{\sqrt{4\pi}\hbar} })\mbp$, 
and define a  dimensionless $s$-wave scattering cross section by
\[\frac{d\til{\s}_{+-}}{d\Omega}=\Lm^{-2}\frac{d{\s}_{+-}}{d\Omega}=\frac{(a/\Lm)^2}{1+4\pi(a/\Lm)^2\til{p}_r^2}\,,
\]
where here and below, we denote dimensionless quantities by a tilde.
With this rescaling, we define dimensionless collision matrix elements by
\[\mathsf{C}_{nm}\equiv{\hbar\over\Lm^5}\til{\mathsf{C}}_{nm}\left({\lm_\z\over\Lm},{a\over\Lm}\right)\,.\]
Here, we expressed the fugacity  $z$=$\z{\rm Li}^{-1}_{3/2}\left(\z\rho{\Lm}^{3}\right)$ as a function of  $\lm_\z$, where we define $\lm_1$=$\lm_B$=$\rho^{-1/3}$  for bosons, $\lm_{-1}$=$\lm_F$=$2\pi(6\pi^2\rho)^{-1/3}$ is the Fermi wavelength, and $\rho$ is the equilibrium density. 
Eq.~\eqref{Cnm} gives the explicit expression for $\til{\mathsf{C}}_{nm}$.  Then, from Eq.~\eqref{L2},  the transport coefficients have the scaling form 
\ben
\hbar\lm_\z L_{\al\be}\equiv\frac{\lm_\z}{\Lambda}\til{L}_{\al\be}\left({\lm_\z\over\Lm},{a\over\Lm}\right)\,,
\label{Lscaling}
\een 
where $\til{L}_{\al\be}\propto\til{\mathsf{C}}_{nm}^{-1}{l}_m{l}_n$.  Then,  we express the parameters above as functions of temperature as
\begin{align*}
 \frac{\lm_\z}{\Lambda}&=y_\z^{-1/3}\sqrt{T\over T_\z}\,,\quad
{a\over\Lm}={a\over\lm_\z}y_\z^{-1/3}\sqrt{T\over T_\z}\,,
\end{align*}
where $T_1$=$T_c$ is the critical temperature for Bose-Einstein condensation given by $\rho\Lm_c^3$=Li$_{3/2}(1)$$\simeq$$2.612$, 
while for fermions, $T_{-1}$=$T_F$ is the Fermi temperature defined by $k_BT_F$=$(2\pi\hbar)^2/2m\lm_F^2$.
It will also be useful to note the relation $ \rho \Lambda ^3$=$x_{\zeta }({T}/{T_{\zeta }})^{-3/2}$, where we define the constants $x_{-1}$=$4/3\sqrt{\pi}$, $y_{-1}$=$\pi^{-3/2}$ and  $y_1$=$x_1$=$2.612$.  

We first define the dimensionless transport coefficients given in terms of dimensionless variables
\begin{align}
\hbar\lm_\z\s_s&={y_{\zeta }}^{-1/3}\pfrac{T}{T_{\zeta }}^{1/2} \tilde{L}_{FF}\,,\nn
\frac{\hbar  \lambda _{\zeta}\kappa_s}{k_B^2 T_{\zeta }}&={y_{\zeta }}^{-1/3}\pfrac{T}{T_{\zeta }}^{3/2}\frac{\det  \tilde{L}}{\tilde{L}_{{FF}}}\,,\nn
{S_s\over k_B}&={\til{L}_{FT}\over\til{L}_{FF}}\,,\quad
 Z_sT=\frac{\til{L}^2_{FT}}{\det\til{L}}\,.\notag
\end{align}
We plot these coefficients Fig.~\ref{BoseCoefs} for bosons, and refer the reader to Ref.~[\onlinecite{wongPRA12}] for the corresponding plots for fermions.
Fig.~\ref{BoseCoefs}~(a) shows the spin conductivity (in color) together with the spin diffusivities (in black), which is the transport coefficient measured in experiments as it determines the spin current driven by spin density gradients via $\mbj_{s}$=$-D_s\bsdel \rho_s$.\cite{sommerNAT11}
It is related to the spin conducitivity by $D_s$=$\s_s/\chi_s$, where $\chi_s$=$\p \rho_s/\p\mu_s$ is the static spin susceptibility and $\mu_s$=$\mu_+$$-$$\mu_-$  is the spin accumulation; expressed in units of $\hbar/m$, it is given by
\[{m\over\hbar}D_s={m\over\hbar}\frac{{{{\sigma}_s }}}{  \chi _s}
=\frac{2 \pi\til{\sigma}_s}{\zeta  \text{Li}_{\frac{1}{2}}(\zeta  z)}\,,\]
where $ \til{\sigma}_s=\hbar\Lm\s_s$. 
The decrease in $\s_s$ and $D_s$ as a function of $T-T_c$ and $a$ is due to the Bose enhancement of scattering.  The spin-heat conductivity at zero current, $\kap_s$, plotted in Fig.~\ref{BoseCoefs}~(b), behaves similarly.  

The spin-Seebeck coefficient and thermospin figure of merit $Z_sT$ are plotted in Fig.~\ref{BoseCoefs} (c) and (d).   For strong scattering ($a/\lm_B$$\geq$1), these coefficients are strongly enhanced near the critical temperature, but at the cost of much shorter spin-heat relaxation lengths $\lm_{\rm st}$\textless1$~\mu$m.  
For weak scattering ($a/\lm_B$$\leq$1), we find $S_s/k_B$$\simeq$0.1 at $(T-T_c)/T_c$=0.01, which is much larger than the case for fermions.\cite{wongPRA12} 
 We also note that for weak scattering a sign change in the spin-Seebeck coefficient as a function of temperature.\cite{wongPRA12}  In contrast, for fermions the spin-Seebeck coefficient changes sign for strong scattering ($a/\lm_F$$\geq$1).  We attribute the sign change to a crossover from particle to hole-dominated transport, as discussed Ref.~[\onlinecite{wongPRA12}].   

For bosons in the degenerate limit, according to he dynamical theory of critical phenomena,\cite{hohenbergRMP77}  transport coefficients exhibit power-law  behavior.  This can be seen by rescaling lengths by the correlation length $\xi$
\ben
\hbar L_{\al\be}\equiv\frac{1}{\xi}\til{L}_{\al\be}\left({\lm\over\xi},{\Lm\over\xi},{a\over\xi}\right)\,,
\label{xiscaling}
\een 
where\cite{pathriaBook11} [See Eq.~\eqref{xi}]
\[\xi={\Lm\over2\pi^{1/2}}\pfrac{-\mu}{ k_BT}^{-1/2}\,.\]
Defining $t$=$({T-T_c})/{T_c}$,  the correlation length diverges as $\xi$$\sim$$ t^{-1}$, see Eq.~\eqref{mu3D}. Since $\til{L}_{\al\be}$ is analytic in $\lm,a$, at degenerate temperatures, a power-law dependence on $\xi$ and $t$ follows from the scaling relation Eq.~\eqref{xiscaling}.   
 {The critical phenomena associated the two-component gas at equal density was studied in Ref.~[\onlinecite{hohenbergRMP77}], where it was called ``the symmetric binary  fluid."}

In the high-temperature limit, the transport coefficients have the scaling form
\[\hbar L_{\al\be}\equiv\frac{1}{\Lm}\til{L}_{\al\be}\left({a\over\Lm}\right)\,,\]
which is the same for bosons and fermions. 
As shown in Eq.~\ref{xihighT}, $\Lm$ is the correlation length in this limit. 
We refer to Fig.~4 of Ref.~[\onlinecite{wongPRA12}] for plots of the transport coefficients in the high-temperature limit as functions of $a/\Lm$.


%

The spin-heat relaxation rate and length,  cf.~Eq.~\eqref{Eq:STrelax1}, in units of ${k_BT_\z/\hbar}$ and in units of $\lm_\z$, respectively,  are given in terms of dimensionless variables by 
\begin{align}
{\hbar\over k_BT_\z\tau_{\rm st}}&={1\over x_\z}\pfrac{T}{T_\z}^{5/2}\frac{\til{\mathsf{C}}_{\rm st}(z,a/\Lm)}{c_p(\z z)}\,,\\
{\lm_{st}\over\lm_\z}&={1\over\lm_\z}\sqrt{\frac{\kappa_s'\tau_{\rm st}}{\rho c_p}}
=y_\z^{1/3}\sqrt{T_\z\over T}\sqrt{\frac{\tilde{L}_{{TT}}}{\til{\mathsf{C}}_{\text{st}}}}\,.
\label{Eq:STrelax}
\end{align}
where we defined
\[{\til{\mathsf{C}}_{\rm st}(z,a/\Lm)}\equiv{\Lambda ^3 \hbar}\left<\pfrac{\e}{ k_BT} \hatC_s\pfrac{\e}{ k_BT}\right>\,,\]
and the heat capacity at constant pressure $c_p$ is given in Eq.~\eqref{cp} and plotted Fig.~\ref{cpfig}. These relaxation coefficients are plotted in Fig.~\ref{STrelax} and their qualitative behavior is discussed in Sec.~\ref{Ts}.  As mentioned before, for bosons the divergence in $\tau_{\rm st}$ stems from the divergence of the heat capacity, plotted Fig.~\ref{cp},  as $T\to T_c$.  For fermions, where the heat capacity remains finite, see Fig.~\ref{cp}(b), the downturn occurs because of Pauli blocking which inhibits scattering.  Thus, we find that the relaxation times diverges at degenerate temperatures, so that the spin-heat accumulation is in principle well-defined for degenerate gases.   

\section{Conclusion and outlook}

In summary, we have developed the theory of coupled spin and heat transport in ultracold atomic gases at degenerate temperatures, including quantum effects due to Bose and Fermi statistics and quantum mechanical scattering.
Using a perturbative solution to the Boltzmann equations that explicitly respects the Onsager reciprocity principle, we computed the  spin-heat transport and relaxation coefficients.  We find a divergence in the spin-heat relaxation times at degenerate temperatures, and that the spin-heat relaxation lengths can be of the order of mm's.  This raises the hope that the spin to heat conversion studied in this work, Ref.~\onlinecite{wongPRL12} and Ref.~\onlinecite{wongPRA12} can be achieved in ultracold atom experiments.  Specifically, using the spin-Seebeck coupling, pure spin-heat currents and resulting spin-dependent heating can be generated by spin-forces even in a gas with equal densities of spin up and spin down particles. 

In this work, we have only touched upon the spin hydrodynamics of two-component gases, which is rich and complex even in the classical regime, and experimental efforts in this direction have only recently begun.  We expect much more interesting and possibly useful physics to emerge in this subject,
and hope this work will motivate further experimental efforts in studying thermospin effects in ultracold atomic gases.

This work was supported by the Stichting voor Fundamenteel Onderzoek der Materie (FOM), by the
European Research Council (ERC) under the Seventh Framework Program (FP7), and is part of the
D-ITP Consortium, a program of the Netherlands Organisation for Scientific Research (NWO) that is funded by the Dutch Ministry of Education, Culture and Science (OCW).

\appendix
 
\section{Thermodynamic properties\label{sec:thermo}}
In this appendix, we summarize the equilibrium properties of noninteracting degenerate gases.  The local density,\cite{Note10} energy density,  pressure, and entropy per particle are given by 
\begin{align}
\coltwo{\rho_\s(\mu_\s,T_\s)}{e_\s(\mu_\s,T_\s)}
&=\int {d^3p\over(2\pi\hbar)^3}\,\coltwo{1}{\e_\mbp}{1 \over z_\s^{-1}e^{\e_\mbp/k_BT_\s}-\z}\nn=&{\z\over\Lm^3}\coltwo{{\rm Li}_{3/2}(\z z_\s)}{(3k_BT/2){\rm Li}_{5/2}(\z z)}
\,,\nn
{p_\s(\mu_\s,T_\s)}&=k_B\int {d^3p\over(2\pi)^3}\,\ln( 1-e^{(\e_\mbp-\mu_\s)/k_BT_\s})\nn
&={\rm Li}_{5/2}(z){k_BT\over\Lm^3}\,,\nn
\frac{s_{\zeta }(z)}{k_B}&={5\over2}\frac{ \text{Li}_{{5}/{2}}(\zeta z)}{\text{Li}_{{3}/{2}}(\zeta z)}-\ln z\,,
\label{thermo}
\end{align}
where $\Lambda=\hbar\sqrt{2\pi/mk_BT}$ is the thermal deBroglie wavelength and $z_\s=e^{\mu_\s/k_BT_\s}$ the fugacity.   The pressure is related to the energy density by $p_\s={2}e_\s/3$, giving the equation of state
\[p_\s(\rho_\s,T_\s)=\rho_\s k_BT_\s \frac{{\rm Li}_{5/2}(\zeta z_\s)}{{\rm Li}_{3/2}(\zeta z_\s)}\,,\]
where in the above the chemical potential is meant to be expressed in terms of the density by 
\[\mu_\s(\rho_\s,T_\s)=\z k_BT_\s {{\rm Li}^{-1}_{3/2}(\z\rho_\s\Lm^3)}\,.\]
The energy and enthalpy per particle, $u_\s\equiv{e_\s}/{\rho_\s}$ and $w_\s\equiv u_\s+p_\s/\rho_\s={5\over3}u_\s$, respectively, are given by 
\[\coltwo{u_\s(\mu_\s,T_\s)}{w_\s(\mu_\s,T_\s)}=\coltwo{3/2}{5/2}k_BT_\s\frac{{\rm Li}_{5/2}(\zeta z_\s)}{{\rm Li}_{3/2}(\zeta z_\s)}\,. \]
\begin{figure}[t]
\begin{center}
\includegraphics[width=.7\linewidth]{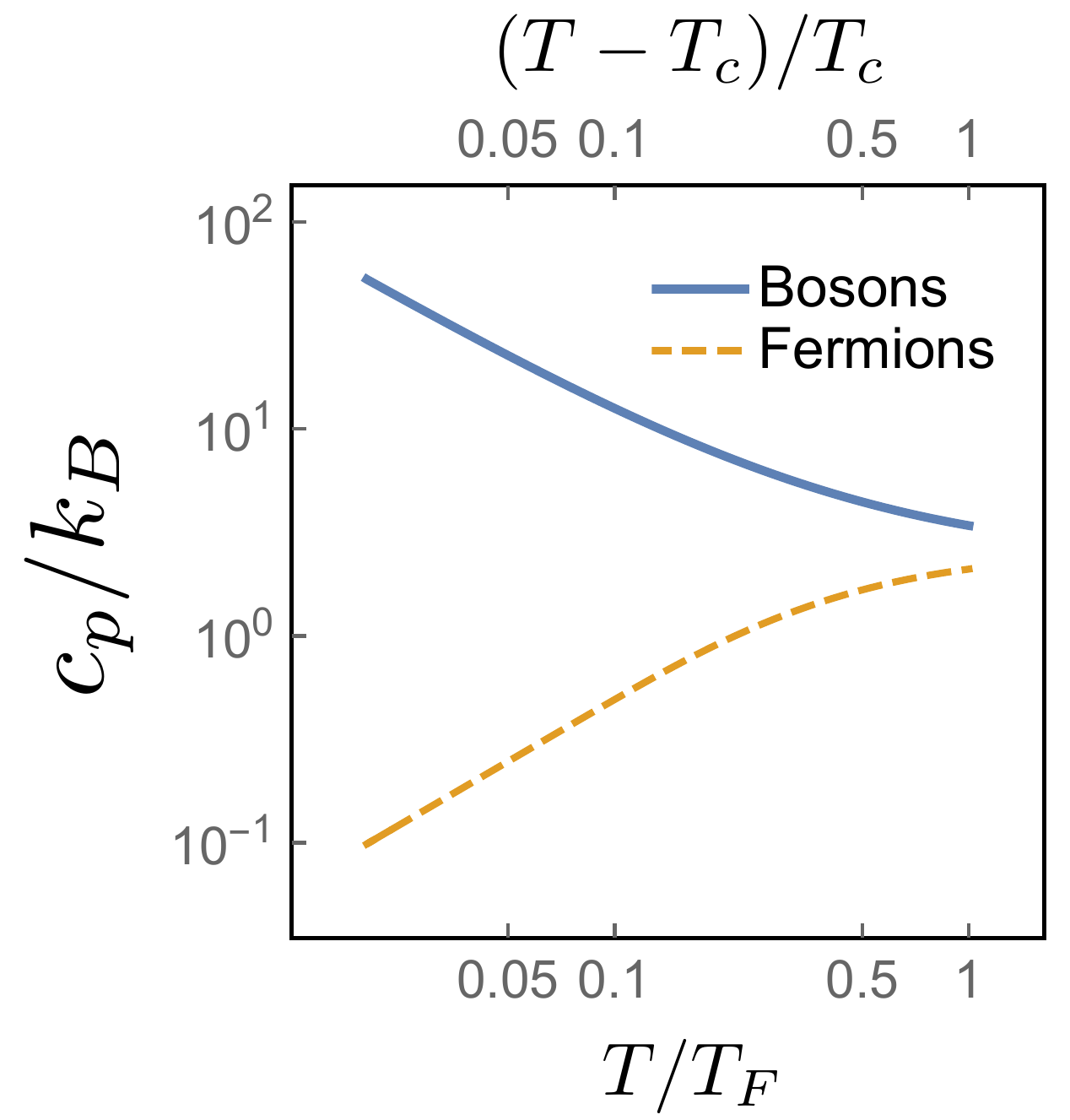}
\caption{The specific heat capacity at constant pressure in units of $k_B$, $c_p/k_B$, of the ideal Bose and Fermi gas, as a function of $(T-T_c)/T_c$ and $T/T_F$, respectively}
\label{cpfig}
\end{center}
\end{figure}
The polylogarithms arise through the integrals
\[\int_0^\infty dx {x^{s-1}\over z^{-1}e^x-\zeta}=\zeta\Gamma_s{\rm Li}_s(\zeta z)\,,\]
where
\[\Gamma_s=\int_0^\infty\,dx\,x^{s-1}e^{-x}\]
 is the gamma function.  They have the series expansion
\[{\rm Li}_s(z)=\sum_{n=1}^{\infty}{z^n\over n^s}=z+{z^2\over2^s}+\ldots\,,\] 
and satisfy the recursion relations 
\[z\p_z {\rm Li}_s={\rm Li}_{s-1}.\] 
The specific heat capacity at constant pressure  can be expressed in terms of the enthalpy change as, 
\[c_{p}=T\pfrac{\p s}{ \p T}_p=\pfrac{\p w}{\p T}_{p}\,.\]
Using the formulae in Eq.~\eqref{thermo} and the identity $(\p\mu/\p T)_p=-s$, we find\cite{pathriaBook11}
\ben
 \frac{c_p(\z z)}{k_B}={25\over4}\frac{\text{Li}_{\frac{1}{2}} (\zeta  z)\text{Li}^2_{\frac{5}{2}} (\zeta  z)}{\text{Li}^3_{\frac{3}{2}} (\zeta  z)}-{15\over4}\frac{\text{Li}_{\frac{5}{2}} (\zeta  z)}{\text{Li}_{\frac{3}{2}} (\zeta  z)}\,.
 \label{cp}
\een
The heat capacity for bosons and fermions are plotted in Fig.~\ref{cpfig} (a) and (b), respectively.  For bosons, $c_p$  diverges as one approaches the Bose-Einstein phase transition.

\begin{figure}[t]
\includegraphics[width=\linewidth]{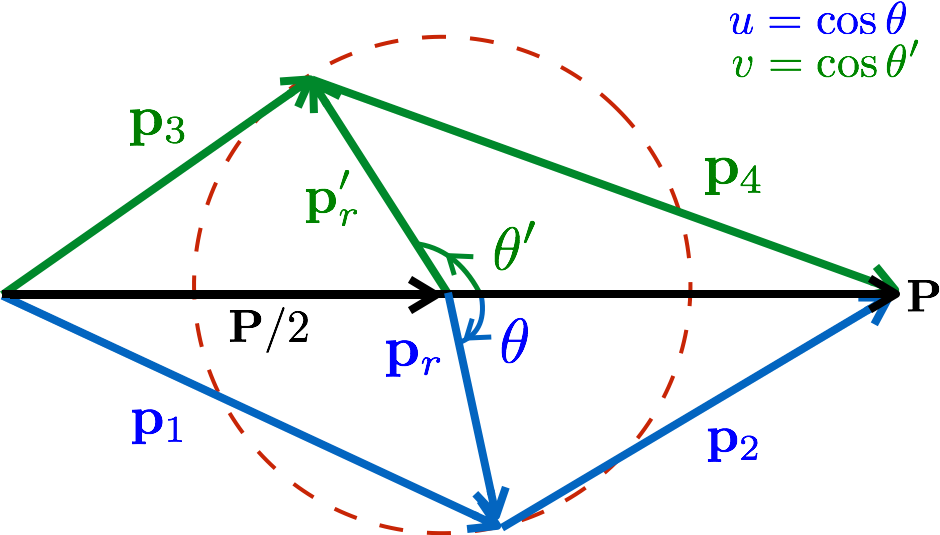}
\caption{(Color online) Coordinates for the two-body scattering angles.  The red dashed circle indicate the constraint $|\mbp_r|=|\mbp_r|'$ due to energy conservation.}
\label{angles}
\end{figure}

We next derive the correlation length in the Bose gas and the classical gas from the local equilibrium distribution for the one-component gas.\cite{Note11}   The one-particle correlation function is related to the semiclassical distribution function  by \cite{Note12}
 \begin{align}
G(\mbr) &=\int {d\mbp\over(2\pi\hbar)^3}\frac{e^{i\mbp\cdot\mbr/\hbar}}{z^{-1}e^{\e_\mbp/k_BT}-\zeta}\,.
\end{align}
In the high-temperature limit $z\ll1$, one finds
\ben
G(r)={z\over\Lambda^3} e^{-{\pi  r^2}/{\Lambda^2}}\,,
\label{xihighT}
\een
where $r=|\mbr|$ so that $\Lm$ is the correlation length for classical thermal fluctuations. 

In the  limit of degenerate temperatures, as $T$$\to$$T_c$, for  $r$$\ll$$\xi$, one finds\cite{pathriaBook11}
\ben
G(r)\approx\frac{e^{-r/\xi}}{\Lm^2 }\,,\quad \xi={\Lm\over2\pi^{1/2}}\pfrac{-\mu}{ k_BT}^{-1/2}\,,
\label{xi}
\een
thus $\xi$ is the correlation length.  To show that it diverges as $T\to T_c^+$, consider the asymptotic expansion in the limit $T\to T_c^+$ 
\ben
{\rm Li}_{d/2}(z)\sim\zeta_{d/2}-\Big|\G_\frac{2-d}{2}\Big |\pfrac{-\mu}{ k_BT}^{d-2\over2}\,, 
\label{asymLi}
\een
where $\zeta_{d/2}$ is the Riemann zeta function, i.e., $\zeta_{3/2}$=$2.612$, $\zeta_{5/2}$=$1.314$.  Then, solving for $\mu$ from  $\rho\Lm^3$=${\rm Li}_{3/2}(z)$, one finds 
\ben
\frac{-\mu}{ k_BT}\to\pfrac{(3/2)\zeta_{3/2}}{\G_{1/2}}^2t^2 \,,\quad t\equiv\frac{T-T_c}{T_c}\,.
\label{mu3D}
\een
where $t$ is the reduced temperature.
From Eq.~\eqref{mu3D} and \eqref{xi}, we have $\xi$$\sim$$ t^{-1}$.  Substituting Eq.~\eqref{mu3D}  into \eqref{cp} gives the critical exponent for the power-law dependence of $c_p$ on $t$.  A similar procedure can be done to extract the critical exponent for all the bosonic transport coefficients calculated in this paper.

\section{Evaluation of collision matrix elements\label{collisionMat}}

To evaluate the collision matrix elements, in Eq.~\eqref{cmat}, we go to center of mass coordinates [See Fig.~\ref{angles}]
\ben
\mbp_{1,2}={\mbP\over2}\pm \mbp_r\,,\quad \mbp_{3,4}={\mbP'\over2}\pm\mbp'_r\,, 
\label{cm}
\een
Energy and momentum conservation gives $p_r$$\equiv$$|\mbp_r|$=$|\mbp_r'|$ and $\mbP$=$\mbP'$.  Furthermore, we use the Hermitian property of the collision operator\cite{landauPK} to express collision matrix elements in a symmetric form 
\begin{widetext}
\begin{align}
\mathsf{C}_{mn}&={1 \over 2mk_BT}\int{d\mbP d\mbp_r\over(2\pi\hbar)^6} p_r
 \int d\W_rd\Omega_r'\, F(z;p_r,P, \bfhat{P}\cdot\bfhat{p_r},\bfhat{P}\cdot\bfhat{p_r} ' )
\sum_\s\left\{{d\s_{+\s}\over d\Omega}\D_{+\s}\left[\pfrac{\e_\mbp}{ k_BT}^n\mbp\right]\cdot\D_{+\s}\left[\pfrac{\e_\mbp}{ k_BT}^m\mbp\right]\right\}\,,
\end{align}
where we define a function of the phase space occupation functions with energy and momentum conservation enforced,
\begin{align}
F&(z;p_r,P, \bfhat{P}\cdot\bfhat{p_r},\bfhat{P}\cdot\bfhat{p_r} ' ) \equiv f^0_1f^0_2(1+\z f^0_3)(1+\z f^0_4)
=\frac{ z^{2} e^{(\e_3+\e_4)/k_BT}}{(e^{\e_1/k_BT}-\zeta z){(e^{\e_2/k_BT}-\zeta z)}{(e^{\e_3/k_BT}-\zeta z)}{(e^{\e_4/k_BT}-\zeta z)}} \,,\nn
{\e_{1,2}\over k_BT}&=\frac{P^2/4+p_r^2\pm P p_r \bfhat{P}\cdot\bfhat{p_r} }{2mk_BT}\,,\quad
{\e_{3,4}\over k_BT}=\frac{P^2/4+p_r^2\pm Pp_r \bfhat{P}\cdot\bfhat{p_r} '}{2mk_BT}\,,
\label{phase}
\end{align}
where the fugacity $z=e^{\mu/k_BT}$  is determined by the density and temperature.  To factor out the dimensionful quantities, here and below, we define dimensionless momenta by the rescaling,
\[{\mbp}\to{\sqrt{2mk_BT}}\mbp=\frac{\sqrt{ 4\pi}\hbar }{\Lambda }\mbp\,.\]
Letting $(\theta,\varphi)$ and $(\theta',\varphi')$ be the spherical angles of $\mbp_r$ and $\mbp_r'$, taking the $z$-axis to lie along $\mbP$ (See Fig.~\ref{angles}), and defining $u$=$\cos\theta$, $v$=$\cos\theta'$, we have
\begin{align}
\mathsf{C}_{nm}
&=\frac{2\hbar}{ \pi ^{5/2} \Lambda ^7}\int_0^\infty  4\pi P^2 dP\int_0^\infty d p_r\,p_r^3
\sum_{\s=\pm} \Lm^2 I_{nm}^{+\s}(z,a/\Lm;P,p_r)
\equiv\frac{\hbar}{\Lambda^5}\til{\mathsf{C}}_{nm}\left(z,{a\over\Lm}\right) \,,
\label{Cnm}\\
 I_{nm}^{+\s}(z,a/\Lm;P,p_r)&=(2\pi)^2 \int_{-1}^1\int_{-1}^{1} dudv
  {d\til{\s}_{+\s}(p_r,a/\Lm)\over d\Omega}F(z;u,v,P,p_r) D^{+\s}_{nm}(u,v,P,p_r)\,,\nn
\frac{d\til{\s}_{\s\tau}}{d\Omega}&=\Lm^{-2}\frac{d{\s}_{\s\tau}}{d\Omega}
\,\quad
D^{+\s}_{nm}=\<\D_{+\s} \left(p^n\mbp\right)\cdot\D_{+\s}(p^m\mbp)\>\,,\label{Imn}
\end{align}
where we denote by brackets the angular average
\[ \langle\ldots\rangle\equiv\int {d\varphi d\varphi'\over(2\pi)^2} \int dp_r'\de (p_r-p_r')\ldots\,,\]
and we defined a dimensionless collision integral matrix elements and differential cross section,  $\til{\mathsf{C}}_{nm}$ and ${d\til{\s}_{\s\tau}}/{d\Omega}$, respectively.   We consider in this paper only spherically symmetric scattering cross sections that do not depend on  the azimuthal angles. The angular integrations over $u,v$ can be evaluated analytically, and the integrals of $P$ and $p_r$ can be evaluated numerically.  

In the high-temperature limit, $z=\rho\Lm^3\ll1$, the quantum statistical factors can be neglected, and the distribution function has the Maxwell-Boltzmann form.  The collision integrals then read,
\begin{align}
 I_{nm}^{+\s}&(z,a/\Lm;P,p_r)\to(2\pi)^2\rho_+\rho_\s e^{-P^2/2-2p_r^2} \int_{-1}^1\int_{-1}^{1} dudv\, {d\til{\s}_{+\s}(p_r,a/\Lm)\over d\Omega} D^{+\s}_{nm}(u,v,P,p_r)\,,
 \label{Imncl}
\end{align}
and can be expressed in terms of incomplete Gamma functions.

Similarly, for the computation of the spin-heat relaxation rate, we encounter the following integral
\begin{align}
\left<\pfrac{\e_\mbp}{k_BT} \hatC_s\pfrac{\e_\mbp}{k_BT}\right>
&={1\over2\pi^{7/2}{\Lambda^5 \hbar }}\int_0^\infty  4\pi P^2 dP\int_0^\infty d p_r\,p_r^3
\Lm^2 I^{+-}_{\rm st}(z,a/\Lm;P,p_r)
\equiv\frac{\til{\mathsf{C}}_{\rm st}(z,a/\Lm)}{\Lambda ^3 \hbar}\nn
I^{+-}_{\rm st}(z,a/\Lm;P,p_r)&=(2\pi)^2\int_{-1}^1\int_{-1}^{1} dudv
{d\til{\s}_{+-}(p_r,a/\Lm)\over d\Omega}F(z;u,v,P,p_r)D_{\rm st}(P,p_r,u,v)\,,\nn
D_{\rm st}(P,p_r,u,v)&=\<\D_{+-} (\e_\mbp)^2\>=4P^2p_r^2(v-u)^2\,.
\label{Cst}
\end{align}
The angular factors are calculated in the next section.  In three dimensions, the angular integrations can be done analytically.  The simplest one is given by
  \begin{align}
 &I_{00}^{+-}(z;P,p_r)={d\til{\s}_{+\s}(p_r,a/\Lm)\over d\Omega}{512\sqrt{\pi }\over\frac{2}{ \pi ^{5/2}} 4\pi P^2}\frac{z^2e^{\frac{P^2}{2}+2 p_r^2} }{\left(e^{\frac{P^2}{2}+2 p_r^2}-z^2\right)^2}
 \ln\left[e^{-\frac{1}{2}\left(p_1^2-p_2^2\right)}\frac{e^{p_1^2}-z}{e^{p_2^2}-z}\right] \ln  \left[e^{-\left(p_1^2-p_2^2\right)}\left(\frac{e^{p_1^2}- z}{e^{p_2^2}- z}\right)^2\right]\,,
\end{align}
where $p_{1,2}$ are to be expressed in terms of center of mass coordinates [cf. Eq.~\eqref{cm}].  
In previous work\cite{drielPRL10} on spin drag, this integrand was written in terms of susceptibilities as function of momentum transfer.

\subsection{Angular integrations \label{Dnm}}

We show some details of the computation of the angular integral Eq.~\eqref{Cnm}, which we reproduce here for convenience,
\begin{align}
 I_{nm}^{+\s}(z;P,p_r)&=\int_{-1}^1\int_{-1}^{1} (2\pi)^2dudvF(z;u,v,P,p_r) D^{+\s}_{nm}(u,v,P,p_r)\,.
\end{align}
\end{widetext}

We first compute $D^{+\s}_{nm}(u,v,P,p_r)$.  Recalling that energy and momentum conservation  results in  $\D_{++}(\mbp)=0$ and $\D_{++}(\e_\mbp)=0$,  $I_{00}$ and $I_{01}$ depends only on inter-spin scattering, so we only need to compute the following  
\begin{align}
&\langle\D^2_{+-}(\e_\mbp)\rangle\,,\,\,\langle\D^2_{+-}(\mbp)\rangle\,,\langle\D_{+-}(\mbp)\cdot\D_{+-}(p^2\mbp)\rangle\,,\nn
&\langle {\D}_+^2(p^2\mbp)\rangle\,,\,\,\langle{\D}_{+-}^2(p^2\mbp)\rangle\,,
\end{align}
where we recall $\langle\ldots\rangle\equiv\int {d\varphi d\varphi'\over(2\pi)^2} \int dp_r'\de (p_r-p_r')\ldots$.  Taking $p_r=p_r'$ ahead of time, we find
\begin{align}
\D_{+-}(\mbp)&=2\mbp_r'-2\mbp_r\,,\quad \D_{+-}(\e_\mbp)=2\mbP\cdot(\mbp_r'-\mbp_r)\nn
\D_{++}(p^2\mbp)&=2(\mbP\cdot\mbp_r')\mbp_r'-2(\mbP\cdot\mbp_r)\mbp_r\label{delta1}\\
{\D}_{+-}(p^2\mbp)&=\mbP\cdot(\mbp_r'-\mbp_r) \mbP+(P^2/2+2p_r^2)(\mbp_r'-\mbp_r)\,,\nonumber
\end{align}
hence
\begin{align}
\D^2_{+-}(\mbp)&=8p_r^2(1-\hat{\mbp}_r\cdot\hat{\mbp}_r')\,,\label{delta2}\\
\D^2_{+-}(\e_\mbp)&=4P^2p_r^2(v-u)^2\,,\nn
\D_{+-}(\mbp)\cdot\D_{+-}(p^2\mbp)
&=2P^2p_r^2(u-v)^2\,,\nn
&\quad+2p_r^2(P^2+4p_r^2)(1-\bfhat{p}_r\cdot\bfhat{p}_r')\,,\nn
\D_{++}^2(p^2\mbp)&=P^2p_r^4[4(u^2+v^2)-8uv\bfhat{p}_r\cdot\bfhat{p}_r']\,,\nn
{\D}_{+-}^2(p^2\mbp)&=P^2p_r^2 (3P^2+4p_r^2)(u-v)^2\nn
&\quad+2p_r^2(P^2/2+2p_r^2)^2(1-\bfhat{p}_r\cdot\bfhat{p}_r')\,.\nonumber
\end{align}
In 3D, the average over azimuthal angles  can be done using the identity 
\[\mbp_r\cdot\mbp_r'=\sin\theta\sin\theta'\cos(\varphi-\varphi')+\cos\theta\cos\theta'\,,\] 
and since $\langle\cos(\varphi-\varphi')\rangle=0$, we have $\langle\mbp_r\cdot\mbp_r'\rangle=uv$, hence
\begin{align}
D^{++}_{00}&=\langle\D^2_{++}(\mbp)\rangle=0\,,\nn  
D^{+-}_{00}&=\langle\D^2_{+-}(\mbp)\rangle=8p_r^2(1-uv)\,, \nn
D^{+-}_{\rm st}&=\<\D^2_{+-}(\e_\mbp)\>=4P^2p_r^2(v-u)^2\,,\nn
D^{++}_{01}&=0\,,\nn
D^{+-}_{01}&=\langle\D_{+-}(\mbp)\cdot\D_{+-}(p^2\mbp)\rangle=2p_r^2[P^2(u-v)^2\nn
&\quad+(P^2+4p_r^2)(1-uv))]\,,\nn
D^{++}_{11}&=\langle {\D}_{++}^2(p^2\mbp)\rangle=P^2p_r^4[4(u^2+v^2)-8(uv)^2]\,,\nn
D^{+-}_{11}&=\langle{\D}_{+-}^2(p^2\mbp)\rangle =p_r^2[P^2  (3P^2+4p_r^2)(u-v)^2\nn
&+2(P^2/2+2p_r^2)^2(1-uv)]\,.
\end{align}

\section{Comparison with spin drag relaxation time\label{spindrag}}
In this section, we show that the leading term in the solution given in Sec.~\ref{moment} is consistent with the spin-drag relaxation time $\tau_{\rm sd}$, defined by  $\s_s$=$\rho\tau_{\rm sd}/m$,  which has been computed in the literature in the absence of spin-heat currents.\cite{duinePRL09,drielPRL10}  For this purpose, we write the the expansion coefficients as
\ben
\begin{pmatrix}c_0^{(F)}\\c_1^{(F)}\end{pmatrix}
=\frac{3\rho}{(1-\mathsf{C}_{01}^2/\mathsf{C}_{00}\mathsf{C}_{11})}
\begin{pmatrix}
{1\over \mathsf{C}_{00}}-{w\over k_BT}\frac{ \mathsf{C}_{01}}{\mathsf{C}_{00}\mathsf{C}_{11}}\\\\
\frac{w }{k_BT\mathsf{C}_{11}}-{\mathsf{C}_{01}\over \mathsf{C}_{00} \mathsf{C}_{11}}
\end{pmatrix}\,.
\een
\vspace{.1cm}

From power counting, $\mathsf{C}_{nm}$$\propto$$ \int dp\,  p^{7+2n+2m}$, we expect that $\mathsf{C}_{00}$\textless$\mathsf{C}_{01}$\textless$\mathsf{C}_{11}$,  hence $\mathsf{C}_{01}/\mathsf{C}_{00}\mathsf{C}_{11}$$\ll$1 and $\mathsf{C}_{00}/\mathsf{C}_{11}$$\ll$1.  Thus, to leading order in  the ratios $\mathsf{C}_{01}/\mathsf{C}_{00}\mathsf{C}_{11}$ and $\mathsf{C}_{00}/\mathsf{C}_{11}$, the spin conductivity is given by 
\[\s_s=L_{FF}=\rho c_0^{(F)}={3\rho^2\over \mathsf{C}_{00}}\,,\] 
for which,
\ben
{1\over \tau_{\rm sd}}={1\over mc_0^{(F)}}={\mathsf{C}_{00}\over 3m\rho}=\frac{\langle\mbv_\mbp\otimes\hat{C}_s\mbv_\mbp\rangle}{\langle\mbv_\mbp\otimes\mbv_\mbp\rangle}\,.
\een
This expression in terms of inner product is consistent with that of Refs.~[\onlinecite{sommerNAT11,bruunNJP11}]. 

This leading-order solution, given by ${\phi}_{\mbp s}=\tau_{\rm sd}\mbv_p\cdot\mbf{F}_s$, describes a uniform shift of the equilibrium  distributions of the spin up and down particles in opposite directions,  resulting in a spin current.  To this order, the spin conductivity is determined by the viscosity between up and down atoms that arises from inter-spin scattering, hence the name spin drag.\cite{Note13}  In contrast, the spin-heat conductivity, which has dependence on intra-spin scattering, is finite even in the absence of inter-spin scattering.

The second-order solution which we have included in this paper represents a distortion of the local distribution and is necessary to capture coupled spin and heat flows because the energy current carried by the leading-order solution is subtracted in the definition of the heat current, cf.~Eq.~\eqref{heat}.

%


\vspace{.1cm}

\begin{thebibliography}{45}%
\makeatletter
\providecommand \@ifxundefined [1]{%
 \@ifx{#1\undefined}
}%
\providecommand \@ifnum [1]{%
 \ifnum #1\expandafter \@firstoftwo
 \else \expandafter \@secondoftwo
 \fi
}%
\providecommand \@ifx [1]{%
 \ifx #1\expandafter \@firstoftwo
 \else \expandafter \@secondoftwo
 \fi
}%
\providecommand \natexlab [1]{#1}%
\providecommand \enquote  [1]{``#1''}%
\providecommand \bibnamefont  [1]{#1}%
\providecommand \bibfnamefont [1]{#1}%
\providecommand \citenamefont [1]{#1}%
\providecommand \href@noop [0]{\@secondoftwo}%
\providecommand \href [0]{\begingroup \@sanitize@url \@href}%
\providecommand \@href[1]{\@@startlink{#1}\@@href}%
\providecommand \@@href[1]{\endgroup#1\@@endlink}%
\providecommand \@sanitize@url [0]{\catcode `\\12\catcode `\$12\catcode
  `\&12\catcode `\#12\catcode `\^12\catcode `\_12\catcode `\%12\relax}%
\providecommand \@@startlink[1]{}%
\providecommand \@@endlink[0]{}%
\providecommand \url  [0]{\begingroup\@sanitize@url \@url }%
\providecommand \@url [1]{\endgroup\@href {#1}{\urlprefix }}%
\providecommand \urlprefix  [0]{URL }%
\providecommand \Eprint [0]{\href }%
\providecommand \doibase [0]{http://dx.doi.org/}%
\providecommand \selectlanguage [0]{\@gobble}%
\providecommand \bibinfo  [0]{\@secondoftwo}%
\providecommand \bibfield  [0]{\@secondoftwo}%
\providecommand \translation [1]{[#1]}%
\providecommand \BibitemOpen [0]{}%
\providecommand \bibitemStop [0]{}%
\providecommand \bibitemNoStop [0]{.\EOS\space}%
\providecommand \EOS [0]{\spacefactor3000\relax}%
\providecommand \BibitemShut  [1]{\csname bibitem#1\endcsname}%
\let\auto@bib@innerbib\@empty
\bibitem [{\citenamefont {Bauer}\ \emph {et~al.}(2012)\citenamefont {Bauer},
  \citenamefont {Saitoh},\ and\ \citenamefont {van Wees}}]{bauerNAT12}%
  \BibitemOpen
  \bibfield  {author} {\bibinfo {author} {\bibfnamefont {G.~E.~W.}\
  \bibnamefont {Bauer}}, \bibinfo {author} {\bibfnamefont {E.}~\bibnamefont
  {Saitoh}}, \ and\ \bibinfo {author} {\bibfnamefont {B.~J.}\ \bibnamefont {van
  Wees}},\ }\href@noop {} {\bibfield  {journal} {\bibinfo  {journal} {Nat
  Mater}\ }\textbf {\bibinfo {volume} {11}},\ \bibinfo {pages} {391} (\bibinfo
  {year} {2012})}\BibitemShut {NoStop}%
\bibitem [{\citenamefont {Brantut}\ \emph {et~al.}(2013)\citenamefont
  {Brantut}, \citenamefont {Grenier}, \citenamefont {Meineke}, \citenamefont
  {Stadler}, \citenamefont {Krinner}, \citenamefont {Kollath}, \citenamefont
  {Esslinger},\ and\ \citenamefont {Georges}}]{brantutSCI13}%
  \BibitemOpen
  \bibfield  {author} {\bibinfo {author} {\bibfnamefont {J.-P.}\ \bibnamefont
  {Brantut}}, \bibinfo {author} {\bibfnamefont {C.}~\bibnamefont {Grenier}},
  \bibinfo {author} {\bibfnamefont {J.}~\bibnamefont {Meineke}}, \bibinfo
  {author} {\bibfnamefont {D.}~\bibnamefont {Stadler}}, \bibinfo {author}
  {\bibfnamefont {S.}~\bibnamefont {Krinner}}, \bibinfo {author} {\bibfnamefont
  {C.}~\bibnamefont {Kollath}}, \bibinfo {author} {\bibfnamefont
  {T.}~\bibnamefont {Esslinger}}, \ and\ \bibinfo {author} {\bibfnamefont
  {A.}~\bibnamefont {Georges}},\ }\href {\doibase 10.1126/science.1242308}
  {\bibfield  {journal} {\bibinfo  {journal} {Science}\ }\textbf {\bibinfo
  {volume} {342}},\ \bibinfo {pages} {713} (\bibinfo {year}
  {2013})}\BibitemShut {NoStop}%
\bibitem [{\citenamefont {Grenier}\ \emph {et~al.}(2012)\citenamefont
  {Grenier}, \citenamefont {Kollath},\ and\ \citenamefont
  {Georges}}]{grenierCM12}%
  \BibitemOpen
  \bibfield  {author} {\bibinfo {author} {\bibfnamefont {C.}~\bibnamefont
  {Grenier}}, \bibinfo {author} {\bibfnamefont {C.}~\bibnamefont {Kollath}}, \
  and\ \bibinfo {author} {\bibfnamefont {A.}~\bibnamefont {Georges}},\
  }\href@noop {} \ \Eprint
  {http://arxiv.org/abs/1209.3942} {arXiv: 1209.3942} \BibitemShut {NoStop}%
\bibitem{ranconCM13}
A.~{Rancon}, C.~{Chin}, and K.~{Levin}, {arXiv: 1311.0769}.
\bibitem [{\citenamefont {{Hazlett}}\ \emph {et~al.}(2013)\citenamefont
  {{Hazlett}}, \citenamefont {{Ha}},\ and\ \citenamefont
  {{Chin}}}]{HazlettCM13}%
  \BibitemOpen
  \bibfield  {author} {\bibinfo {author} {\bibfnamefont {E.~L.}\ \bibnamefont
  {{Hazlett}}}, \bibinfo {author} {\bibfnamefont {L.-C.}\ \bibnamefont {{Ha}}},
  \ and\ \bibinfo {author} {\bibfnamefont {C.}~\bibnamefont {{Chin}}},\
  }\href@noop {} \ \Eprint{http://arxiv.org/abs/1306.4018} {arXiv:1306.4018}\BibitemShut {NoStop}%
\bibitem [{\citenamefont {{Hatami}}\ \emph {et~al.}(2010)\citenamefont
  {{Hatami}}, \citenamefont {{Bauer}}, \citenamefont {{Takahashi}},\ and\
  \citenamefont {{Maekawa}}}]{hatamiSSC10}%
  \BibitemOpen
  \bibfield  {author} {\bibinfo {author} {\bibfnamefont {M.}~\bibnamefont
  {{Hatami}}}, \bibinfo {author} {\bibfnamefont {G.~E.~W.}\ \bibnamefont
  {{Bauer}}}, \bibinfo {author} {\bibfnamefont {S.}~\bibnamefont
  {{Takahashi}}}, \ and\ \bibinfo {author} {\bibfnamefont {S.}~\bibnamefont
  {{Maekawa}}},\ }\href@noop {} {\bibfield  {journal} {\bibinfo  {journal}
  {Solid State Communications}\ }\textbf {\bibinfo {volume} {150}},\ \bibinfo
  {pages} {480} (\bibinfo {year} {2010})}\BibitemShut {NoStop}%
\bibitem [{\citenamefont {{Nunner}}\ and\ \citenamefont {{von
  Oppen}}(2011)}]{nunnerPRB11}%
  \BibitemOpen
  \bibfield  {author} {\bibinfo {author} {\bibfnamefont {T.~S.}\ \bibnamefont
  {{Nunner}}}\ and\ \bibinfo {author} {\bibfnamefont {F.}~\bibnamefont {{von
  Oppen}}},\ }\href {\doibase 10.1103/PhysRevB.84.020405} {\bibfield  {journal}
  {\bibinfo  {journal} {\prb}\ }\textbf {\bibinfo {volume} {84}},\ \bibinfo
  {pages} {020405} (\bibinfo {year} {2011})}\BibitemShut {NoStop}%
      \bibitem [{Note0a()}]{Note0a}%
  \BibitemOpen
  \bibinfo {note}  {For bosons, the pseudospin refers to a sufficiently isolated subset of the atomic hyperfine states, take from, for example, the ground state manifold of $^{87}$Rb discussed in Ref.~[\onlinecite{linNAT11}].}
  \bibitem{linNAT11}
Y.~J. Lin, K.~Jimenez-Garcia, and I.~B. Spielman.
Nature, \textbf{471} 83 (2011).
\bibitem [{\citenamefont {Vichi}\ and\ \citenamefont
  {Stringari}(1999)}]{vichiPRA99}%
  \BibitemOpen
  \bibfield  {author} {\bibinfo {author} {\bibfnamefont {L.}~\bibnamefont
  {Vichi}}\ and\ \bibinfo {author} {\bibfnamefont {S.}~\bibnamefont
  {Stringari}},\ }\href {\doibase 10.1103/PhysRevA.60.4734} {\bibfield
  {journal} {\bibinfo  {journal} {Phys. Rev. A}\ }\textbf {\bibinfo {volume}
  {60}},\ \bibinfo {pages} {4734} (\bibinfo {year} {1999})}\BibitemShut
  {NoStop}%
\bibitem [{\citenamefont {Sommer}\ \emph {et~al.}(2011)\citenamefont {Sommer},
  \citenamefont {Ku}, \citenamefont {Roati},\ and\ \citenamefont
  {Zwierlein}}]{sommerNAT11}%
  \BibitemOpen
  \bibfield  {author} {\bibinfo {author} {\bibfnamefont {A.}~\bibnamefont
  {Sommer}}, \bibinfo {author} {\bibfnamefont {M.}~\bibnamefont {Ku}}, \bibinfo
  {author} {\bibfnamefont {G.}~\bibnamefont {Roati}}, \ and\ \bibinfo {author}
  {\bibfnamefont {M.~W.}\ \bibnamefont {Zwierlein}},\ }\href@noop {} {\bibfield
   {journal} {\bibinfo  {journal} {Nature}\ }\textbf {\bibinfo {volume}
  {472}},\ \bibinfo {pages} {201} (\bibinfo {year} {2011})}\BibitemShut
  {NoStop}%
\bibitem [{\citenamefont {{Koller}}\ \emph {et~al.}(2012)\citenamefont
  {{Koller}}, \citenamefont {{Groot}}, \citenamefont {{Bons}}, \citenamefont
  {{Duine}}, \citenamefont {{Stoof}},\ and\ \citenamefont {{van der
  Straten}}}]{kollerCM12}%
  \BibitemOpen
  \bibfield  {author} {\bibinfo {author} {\bibfnamefont {S.~B.}\ \bibnamefont
  {{Koller}}}, \bibinfo {author} {\bibfnamefont {A.}~\bibnamefont {{Groot}}},
  \bibinfo {author} {\bibfnamefont {P.~C.}\ \bibnamefont {{Bons}}}, \bibinfo
  {author} {\bibfnamefont {R.~A.}\ \bibnamefont {{Duine}}}, \bibinfo {author}
  {\bibfnamefont {H.~T.~C.}\ \bibnamefont {{Stoof}}}, \ and\ \bibinfo {author}
  {\bibfnamefont {P.}~\bibnamefont {{van der Straten}}},\ }\href@noop {} \ \Eprint {http://arxiv.org/abs/1204.6143}{arXiv: 1204.6143} \BibitemShut {NoStop}%
  \bibitem{poliniPRL07}
M.~Polini and G.~Vignale,~{Phys. Rev. Lett.} \textbf{98} 266403 (2007).
\bibitem [{\citenamefont {Weber}\ \emph {et~al.}(2005)\citenamefont {Weber},
  \citenamefont {Gedik}, \citenamefont {Moore}, \citenamefont {Orenstein},
  \citenamefont {Stephens},\ and\ \citenamefont {Awschalom}}]{weberNAT05}%
  \BibitemOpen
  \bibfield  {author} {\bibinfo {author} {\bibfnamefont {C.~P.}\ \bibnamefont
  {Weber}}, \bibinfo {author} {\bibfnamefont {N.}~\bibnamefont {Gedik}},
  \bibinfo {author} {\bibfnamefont {J.~E.}\ \bibnamefont {Moore}}, \bibinfo
  {author} {\bibfnamefont {J.}~\bibnamefont {Orenstein}}, \bibinfo {author}
  {\bibfnamefont {J.}~\bibnamefont {Stephens}}, \ and\ \bibinfo {author}
  {\bibfnamefont {D.~D.}\ \bibnamefont {Awschalom}},\ }\href@noop {} {\bibfield
   {journal} {\bibinfo  {journal} {Nature}\ }\textbf {\bibinfo {volume}
  {437}},\ \bibinfo {pages} {1330} (\bibinfo {year} {2005})}\BibitemShut
  {NoStop}%
\bibitem [{\citenamefont {Ohde}\ \emph {et~al.}(1996)\citenamefont {Ohde},
  \citenamefont {Bonitz}, \citenamefont {Bornath}, \citenamefont {Kremp},\ and\
  \citenamefont {Schlanges}}]{ohdePP96}%
  \BibitemOpen
  \bibfield  {author} {\bibinfo {author} {\bibfnamefont {T.}~\bibnamefont
  {Ohde}}, \bibinfo {author} {\bibfnamefont {M.}~\bibnamefont {Bonitz}},
  \bibinfo {author} {\bibfnamefont {T.}~\bibnamefont {Bornath}}, \bibinfo
  {author} {\bibfnamefont {D.}~\bibnamefont {Kremp}}, \ and\ \bibinfo {author}
  {\bibfnamefont {M.}~\bibnamefont {Schlanges}},\ }\href@noop {} {\bibfield
  {journal} {\bibinfo  {journal} {Physics of Plasmas (1994-present)}\ }\textbf
  {\bibinfo {volume} {3}},\ \bibinfo {pages} {1241} (\bibinfo {year}
  {1996})}\BibitemShut {NoStop}%
\bibitem [{\citenamefont {Sultan}\ \emph {et~al.}(2012)\citenamefont {Sultan},
  \citenamefont {Atxitia}, \citenamefont {Melnikov}, \citenamefont
  {Chubykalo-Fesenko},\ and\ \citenamefont {Bovensiepen}}]{sultanPRB12}%
  \BibitemOpen
  \bibfield  {author} {\bibinfo {author} {\bibfnamefont {M.}~\bibnamefont
  {Sultan}}, \bibinfo {author} {\bibfnamefont {U.}~\bibnamefont {Atxitia}},
  \bibinfo {author} {\bibfnamefont {A.}~\bibnamefont {Melnikov}}, \bibinfo
  {author} {\bibfnamefont {O.}~\bibnamefont {Chubykalo-Fesenko}}, \ and\
  \bibinfo {author} {\bibfnamefont {U.}~\bibnamefont {Bovensiepen}},\ }\href
  {\doibase 10.1103/PhysRevB.85.184407} {\bibfield  {journal} {\bibinfo
  {journal} {Phys. Rev. B}\ }\textbf {\bibinfo {volume} {85}},\ \bibinfo
  {pages} {184407} (\bibinfo {year} {2012})}\BibitemShut {NoStop}%
\bibitem [{\citenamefont {Dejene}\ \emph {et~al.}(2013)\citenamefont {Dejene},
  \citenamefont {Flipse}, \citenamefont {Bauer},\ and\ \citenamefont {van
  Wees}}]{dejeneNATP13}%
  \BibitemOpen
  \bibfield  {author} {\bibinfo {author} {\bibfnamefont {F.~K.}\ \bibnamefont
  {Dejene}}, \bibinfo {author} {\bibfnamefont {J.}~\bibnamefont {Flipse}},
  \bibinfo {author} {\bibfnamefont {G.~E.~W.}\ \bibnamefont {Bauer}}, \ and\
  \bibinfo {author} {\bibfnamefont {B.~J.}\ \bibnamefont {van Wees}},\ }\href
  {http://dx.doi.org/10.1038/nphys2743} {\bibfield  {journal} {\bibinfo
  {journal} {Nat. Phys.}\ }\textbf {\bibinfo {volume} {9}},\ \bibinfo {pages}
  {636} (\bibinfo {year} {2013})}\BibitemShut {NoStop}%
\bibitem{marunPRL14}
J.~Vera-Marun, I.\, J.~van Wees, B.\, and R.~Jansen.
{ Phys. Rev. Lett.} \textbf{112} 056602 (2014).
\bibitem [{Note0()}]{Note0}%
\BibitemOpen
\bibinfo {note} 
 {Our work may also be relevant to the solid-state environment in regards to the contribution to the transport and relaxation coefficients coming from electron-electron interactions.}\BibitemShut {Stop}%
\bibitem [{\citenamefont {Kovalev}\ and\ \citenamefont
  {Tserkovnyak}(2010)}]{kovalevSSC10}%
  \BibitemOpen
  \bibfield  {author} {\bibinfo {author} {\bibfnamefont {A.~A.}\ \bibnamefont
  {Kovalev}}\ and\ \bibinfo {author} {\bibfnamefont {Y.}~\bibnamefont
  {Tserkovnyak}},\ }\href {\doibase DOI: 10.1016/j.ssc.2009.11.012} {\bibfield
  {journal} {\bibinfo  {journal} {Solid State Communications}\ }\textbf
  {\bibinfo {volume} {150}},\ \bibinfo {pages} {500 } (\bibinfo {year}
  {2010})}\BibitemShut {NoStop}%
\bibitem [{\citenamefont {Wong}\ \emph
  {et~al.}(2012{\natexlab{a}})\citenamefont {Wong}, \citenamefont {van Driel},
  \citenamefont {Kittinaradorn}, \citenamefont {Stoof},\ and\ \citenamefont
  {Duine}}]{wongPRL12}%
  \BibitemOpen
  \bibfield  {author} {\bibinfo {author} {\bibfnamefont {C.~H.}\ \bibnamefont
  {Wong}}, \bibinfo {author} {\bibfnamefont {H.~J.}\ \bibnamefont {van Driel}},
  \bibinfo {author} {\bibfnamefont {R.}~\bibnamefont {Kittinaradorn}}, \bibinfo
  {author} {\bibfnamefont {H.~T.~C.}\ \bibnamefont {Stoof}}, \ and\ \bibinfo
  {author} {\bibfnamefont {R.~A.}\ \bibnamefont {Duine}},\ }\href {\doibase
  10.1103/PhysRevLett.108.075301} {\bibfield  {journal} {\bibinfo  {journal}
  {Phys. Rev. Lett.}\ }\textbf {\bibinfo {volume} {108}},\ \bibinfo {pages}
  {075301} (\bibinfo {year} {2012}{\natexlab{a}})}\BibitemShut {NoStop}%
   \bibitem [{Note5()}]{Note5a}%
  \BibitemOpen
  \bibinfo {note} 
 {The s-wave interspin scattering is sufficient to stabilize the spin currents that we consider in this work.  Intraspin relaxation for fermions due to $p$ -wave scattering was considered in Ref.~[\onlinecite{wongPRA12}].}
\bibitem [{\citenamefont {Landau}\ and\ \citenamefont
  {Lifshitz}(1987)}]{landauFM}%
  \BibitemOpen
  \bibfield  {author} {\bibinfo {author} {\bibnamefont {Landau}}\ and\ \bibinfo
  {author} {\bibfnamefont {E.~M.}\ \bibnamefont {Lifshitz}},\ }\href@noop {}
  {\emph {\bibinfo {title} {Fluid Mechanics}}}\ (\bibinfo  {publisher}
  {Pergamon Press},\ \bibinfo {year} {1987})\BibitemShut {NoStop}%
\bibitem [{Note1()}]{Note1}%
  \BibitemOpen
  \bibinfo {note} {For degenerate fermions in solid-state environment, one
  usually subtracts only the chemical potential, related to the enthalpy by
  $w=\mu +Ts$, but the difference is negligible because $s$$\sim$$
  T/T_F$.}\BibitemShut {Stop}%
\bibitem [{Note2()}]{Note2}%
  \BibitemOpen
  \bibinfo {note} {Such a relaxation term is also consistent with the theory of
  quasi-stationary thermodynamics,\cite {LandauBookSP1} which assumes that for
  small deviations from equilibrium, a nonequilibrium quantity such $T_s$
  should decay as $\partial _tT_s$=$-T_s/\tau _{\protect \rm st}$.}\BibitemShut
  {Stop}%
  \bibitem [{\citenamefont {Landau}\ and\ \citenamefont
  {Lifshitz}(1980)}]{LandauBookSP1}%
  \BibitemOpen
  \bibfield  {author} {\bibinfo {author} {\bibfnamefont {L.}~\bibnamefont
  {Landau}}\ and\ \bibinfo {author} {\bibfnamefont {E.}~\bibnamefont
  {Lifshitz}},\ }\href@noop {} {\emph {\bibinfo {title} {Statistical Physics,
  Part 1}}},\ \bibinfo {edition} {3rd}\ ed.\ (\bibinfo  {publisher} {Pergamon
  Press},\ \bibinfo {year} {1980})\BibitemShut {NoStop}%
\bibitem [{Note3()}]{Note3}%
  \BibitemOpen
  \bibinfo {note} {We neglect here the temperature dependence of $\kappa
  _s$}\BibitemShut {NoStop}%
\bibitem [{Note4()}]{Note4}%
  \BibitemOpen
  \bibinfo {note} {{We note that for fermions, a superfluid transition occurs
  at $T_c$$\ll$$T_F$, and that the results reported in this work do not include
  effects related to this transition. Depending on the value of $T_c$, the
  fermi gas may enter a Fermi liquid regime for $T_c$\textless $T$$\ll $$T_F$.
  In this regime, our solution based on Boltzmann equation remain valid
  provided that one uses Fermi-liquid scattering amplitudes in the collision
  integral as shown in Ref.~[\protect \rev@citealpnum
  {bruunNJP11}].}}\BibitemShut {Stop}%
\bibitem [{Note5()}]{Note5}%
  \BibitemOpen
  \bibinfo {note} {The measured values of the transport coefficients should be
  compared with the trap-averaged values, which differ slightly from the
  results presented here.}\BibitemShut {Stop}%
\bibitem [{\citenamefont {Duine}\ and\ \citenamefont
  {Stoof}(2009)}]{duinePRL09}%
  \BibitemOpen
  \bibfield  {author} {\bibinfo {author} {\bibfnamefont {R.~A.}\ \bibnamefont
  {Duine}}\ and\ \bibinfo {author} {\bibfnamefont {H.~T.~C.}\ \bibnamefont
  {Stoof}},\ }\href {\doibase 10.1103/PhysRevLett.103.170401} {\bibfield
  {journal} {\bibinfo  {journal} {Phys. Rev. Lett.}\ }\textbf {\bibinfo
  {volume} {103}},\ \bibinfo {pages} {170401} (\bibinfo {year}
  {2009})}\BibitemShut {NoStop}%
  \bibitem [{\citenamefont {van Driel}\ \emph {et~al.}(2010)\citenamefont {van
  Driel}, \citenamefont {Duine},\ and\ \citenamefont {Stoof}}]{drielPRL10}%
  \BibitemOpen
  \bibfield  {author} {\bibinfo {author} {\bibfnamefont {H.~J.}\ \bibnamefont
  {van Driel}}, \bibinfo {author} {\bibfnamefont {R.~A.}\ \bibnamefont
  {Duine}}, \ and\ \bibinfo {author} {\bibfnamefont {H.~T.~C.}\ \bibnamefont
  {Stoof}},\ }\href {\doibase 10.1103/PhysRevLett.105.155301} {\bibfield
  {journal} {\bibinfo  {journal} {Phys. Rev. Lett.}\ }\textbf {\bibinfo
  {volume} {105}},\ \bibinfo {pages} {155301} (\bibinfo {year}
  {2010})}\BibitemShut {NoStop}%
  \bibitem [{\citenamefont {Kim}\ and\ \citenamefont {Huse}(2012)}]{kimPRA12}%
  \BibitemOpen
  \bibfield  {author} {\bibinfo {author} {\bibfnamefont {H.}~\bibnamefont
  {Kim}}\ and\ \bibinfo {author} {\bibfnamefont {D.~A.}\ \bibnamefont {Huse}},\
  }\href {\doibase 10.1103/PhysRevA.86.053607} {\bibfield  {journal} {\bibinfo
  {journal} {Phys. Rev. A}\ }\textbf {\bibinfo {volume} {86}},\ \bibinfo
  {pages} {053607} (\bibinfo {year} {2012})}\BibitemShut {NoStop}%
\bibitem [{\citenamefont {Chaikin}\ and\ \citenamefont
  {Lubensky}(2000)}]{chaikinBook00}%
  \BibitemOpen
  \bibfield  {author} {\bibinfo {author} {\bibfnamefont {P.}~\bibnamefont
  {Chaikin}}\ and\ \bibinfo {author} {\bibfnamefont {T.}~\bibnamefont
  {Lubensky}},\ }\href {http://books.google.com/books?id=P9YjNjzr9OIC} {\emph
  {\bibinfo {title} {Principles of Condensed Matter Physics}}}\ (\bibinfo
  {publisher} {Cambridge University Press},\ \bibinfo {year}
  {2000})\BibitemShut {NoStop}%
\bibitem [{\citenamefont {Chapman}\ and\ \citenamefont
  {Cowling}(1970)}]{chapmanMTNG70}%
  \BibitemOpen
  \bibfield  {author} {\bibinfo {author} {\bibfnamefont {S.}~\bibnamefont
  {Chapman}}\ and\ \bibinfo {author} {\bibfnamefont {T.}~\bibnamefont
  {Cowling}},\ }\href@noop {} {\emph {\bibinfo {title} {The Mathematical Theory
  of Non-uniform Gases: An Account of the Kinetic Theory of Viscosity, Thermal
  Conduction and Diffusion in Gases}}},\ Cambridge Mathematical Library\
  (\bibinfo  {publisher} {Cambridge University Press},\ \bibinfo {year}
  {1970})\BibitemShut {NoStop}%
\bibitem [{\citenamefont {Pitaevski}\ and\ \citenamefont
  {Lifshitz}(1981)}]{landauPK}%
  \BibitemOpen
  \bibfield  {author} {\bibinfo {author} {\bibfnamefont {L.}~\bibnamefont
  {Pitaevski}}\ and\ \bibinfo {author} {\bibfnamefont {E.~M.}\ \bibnamefont
  {Lifshitz}},\ }\href@noop {} {\emph {\bibinfo {title} {Physical Kinetics}}},\
  \bibinfo {edition} {1st}\ ed.\ (\bibinfo  {publisher} {Pergamon Press},\
  \bibinfo {year} {1981})\BibitemShut {NoStop}%
\bibitem [{Note6()}]{Note6}%
  \BibitemOpen
  \bibinfo {note} {Because in the center-of-mass frame, the reduced mass is
  $m/2$}\BibitemShut {NoStop}%
\bibitem [{\citenamefont {Reichl}(1998)}]{reichlSP98}%
  \BibitemOpen
  \bibfield  {author} {\bibinfo {author} {\bibfnamefont {L.}~\bibnamefont
  {Reichl}},\ }\href@noop {} {\emph {\bibinfo {title} {A Modern Course in
  Statistical Physics}}},\ \bibinfo {edition} {2nd}\ ed.\ (\bibinfo
  {publisher} {John Wiley \& Sons, Inc.},\ \bibinfo {year} {1998})\BibitemShut
  {NoStop}%
\bibitem [{\citenamefont {Mahan}(1997)}]{mahanSSP97}%
  \BibitemOpen
  \bibfield  {author} {\bibinfo {author} {\bibfnamefont {G.}~\bibnamefont
  {Mahan}}\ }(\bibinfo  {publisher} {Academic Press},\ \bibinfo {year} {1997})\
  pp.\ \bibinfo {pages} {81 -- 157}\BibitemShut {NoStop}%
  \bibitem [{\citenamefont {Wong}\ \emph
  {et~al.}(2012{\natexlab{b}})\citenamefont {Wong}, \citenamefont {Stoof},\
  and\ \citenamefont {Duine}}]{wongPRA12}%
  \BibitemOpen
  \bibfield  {author} {\bibinfo {author} {\bibfnamefont {C.~H.}\ \bibnamefont
  {Wong}}, \bibinfo {author} {\bibfnamefont {H.~T.~C.}\ \bibnamefont {Stoof}},
  \ and\ \bibinfo {author} {\bibfnamefont {R.~A.}\ \bibnamefont {Duine}},\
  }\href {\doibase 10.1103/PhysRevA.85.063613} {\bibfield  {journal} {\bibinfo
  {journal} {Phys. Rev. A}\ }\textbf {\bibinfo {volume} {85}},\ \bibinfo
  {pages} {063613} (\bibinfo {year} {2012}{\natexlab{b}})}\BibitemShut
  {NoStop}%
\bibitem [{Note7()}]{Note7}%
  \BibitemOpen
  \bibinfo {note} {In the high-temperature limit, it is more convenient instead
  to expand in terms of the Sonine polynomials which are orthogonal.\cite{chapmanMTNG70,smithTP89,landauPK}}\BibitemShut{NoStop}%
\bibitem [{Note8()}]{Note8}%
  \BibitemOpen
  \bibinfo {note} {Note that in the notation of {Ref.~[\protect \rev@citealpnum
  {wongPRA12}]}, $a_n$=$c_n^{(F)}$ and $b_n$=$c_n^{(T)}$}\BibitemShut{NoStop}%
\bibitem [{Note9()}]{Note9}%
  \BibitemOpen
  \bibinfo {note} {It can be shown that using this method and truncating at a
  finite order always results in transport coefficients which are less than
  their exact values\cite{landauPK,smithTP89}}\BibitemShut{NoStop}%
\bibitem [{\citenamefont {Hohenberg}\ and\ \citenamefont
  {Halperin}(1977)}]{hohenbergRMP77}%
  \BibitemOpen
  \bibfield  {author} {\bibinfo {author} {\bibfnamefont {P.~C.}\ \bibnamefont
  {Hohenberg}}\ and\ \bibinfo {author} {\bibfnamefont {B.~I.}\ \bibnamefont
  {Halperin}},\ }\href {\doibase 10.1103/RevModPhys.49.435} {\bibfield
  {journal} {\bibinfo  {journal} {Rev. Mod. Phys.}\ }\textbf {\bibinfo {volume}
  {49}},\ \bibinfo {pages} {435} (\bibinfo {year} {1977})}\BibitemShut
  {NoStop}%
  \bibitem [{Note10()}]{Note10}%
  \BibitemOpen
  \bibinfo {note} {We are not including the condensate density, which would be
  given by $\rho _c=z/1-z$}\BibitemShut {NoStop}%
\bibitem [{\citenamefont {Pathria}(2011)}]{pathriaBook11}%
  \BibitemOpen
  \bibfield  {author} {\bibinfo {author} {\bibfnamefont {R.}~\bibnamefont
  {Pathria}},\ }\href {http://books.google.com/books?id=KdbJJAXQ-RsC} {\emph
  {\bibinfo {title} {Statistical Mechanics}}}\ (\bibinfo  {publisher} {Elsevier
  Science},\ \bibinfo {year} {2011})\BibitemShut {NoStop}%
\bibitem [{Note11()}]{Note11}%
  \BibitemOpen
  \bibinfo {note} {We neglect spin correlations in this paper.}\BibitemShut
  {Stop}%
\bibitem [{Note12()}]{Note12}%
  \BibitemOpen
  \bibinfo {note} {This follows from the fact that the semiclassical
  distribution function is the Wigner transform of the density matrix. See, for
  example, Ref.~[\protect \rev@citealpnum {naraschewskiPRA99}].}\BibitemShut
  {Stop}%
\bibitem [{\citenamefont {Bruun}(2011)}]{bruunNJP11}%
  \BibitemOpen
  \bibfield  {author} {\bibinfo {author} {\bibfnamefont {G.~M.}\ \bibnamefont
  {Bruun}},\ }\href@noop {} {\bibfield  {journal} {\bibinfo  {journal} {New
  Journal of Physics}\ }\textbf {\bibinfo {volume} {13}},\ \bibinfo {pages}
  {035005} (\bibinfo {year} {2011})}\BibitemShut {NoStop}%
\bibitem [{Note13()}]{Note13}%
  \BibitemOpen
  \bibinfo {note} {However, high order terms do involve intra-spin
  scattering.}\BibitemShut {Stop}%
\bibitem [{\citenamefont {Smith}\ and\ \citenamefont
  {Jensen}(1989)}]{smithTP89}%
  \BibitemOpen
  \bibfield  {author} {\bibinfo {author} {\bibfnamefont {H.}~\bibnamefont
  {Smith}}\ and\ \bibinfo {author} {\bibfnamefont {H.~H.}\ \bibnamefont
  {Jensen}},\ }\href@noop {} {\emph {\bibinfo {title} {Transport Phenomena}}}\
  (\bibinfo  {publisher} {Oxford University Press},\ \bibinfo {year}
  {1989})\BibitemShut {NoStop}%
\bibitem [{\citenamefont {Naraschewski}\ and\ \citenamefont
  {Glauber}(1999)}]{naraschewskiPRA99}%
  \BibitemOpen
  \bibfield  {author} {\bibinfo {author} {\bibfnamefont {M.}~\bibnamefont
  {Naraschewski}}\ and\ \bibinfo {author} {\bibfnamefont {R.~J.}\ \bibnamefont
  {Glauber}},\ }\href {\doibase 10.1103/PhysRevA.59.4595} {\bibfield  {journal}
  {\bibinfo  {journal} {Phys. Rev. A}\ }\textbf {\bibinfo {volume} {59}},\
  \bibinfo {pages} {4595} (\bibinfo {year} {1999})}\BibitemShut {NoStop}%
\end{thebibliography}
\end{document}